\documentclass[preprints,article,accept,pdftex,moreauthors]{Definitions/mdpi} 
\firstpage{1} 
\pubvolume{1}
\issuenum{1}
\articlenumber{0}
\pubyear{2024}
\copyrightyear{2024}
\datereceived{ } 
\daterevised{ } 
\dateaccepted{ } 
\datepublished{ } 
\hreflink{https://doi.org/} 

\usepackage[addedcolor=orange, deletedcolor=red, authormarkup=footnote]{changes}

\newcommand{\eg}{\textit{e.g.}}

\newcommand{\ie}{\textit{i.e.}}
\newcommand{\dd}{\mathrm{d}}

\newcommand{\gsim}{\gtrsim}
\newcommand{\psip}{\psi(2S)}
\newcommand{\jpsi}{J/\psi}
\newcommand{\chic}{\chi_c}
\newcommand{\raa}{R_{\rm AA}}
\newcommand{\npart}{N_{\rm part}}
\newcommand{\pT}{p_{\rm T}}
\newcommand{\Td}{T_{\rm diss}}
\newcommand{\Ncoll}{N_{\rm coll}}

\Title{Charmonium Transport in Heavy-Ion Collisions at the LHC}

\TitleCitation{Title}

\Author{Biaogang Wu $^{1, *}$\orcidA{}, Ralf Rapp $^{1}$\orcidB{}}

\address[1]{%
$^{1}$ \quad Cyclotron Institute and Department of Physics and Astronomy, Texas A$\&$M University, College Station, TX 77843-3366, USA; bgwu@tamu.edu (B.W.); rapp@comp.tamu.edu (R.R.)}

\corres{Correspondence: bgwu@tamu.edu;}

\abstract{
We provide an update on our semi-classical transport approach for quarkonium production in high-energy heavy-ion collisions, focusing on $\jpsi$ and $\psip$ mesons in 5.02 TeV Pb-Pb collisions at the Large Hadron Collider (LHC) at both forward and mid-rapidity. In particular, we employ the most recent charm-production cross sections reported in pp collisions, which are pivotal for the magnitude of the regeneration contribution, and their modifications due to cold-nuclear-matter (CNM) effects. 
Multi-differential observables are calculated in terms of nuclear modification factors as a function of centrality, transverse momentum, and rapidity, including the contributions from feeddown from bottom-hadron decays.
For our predictions for $\psip$ production, the mechanism of sequential regeneration relative to the more strongly bound $\jpsi$ meson plays an important role in interpreting recent ALICE data.
}

\keyword{Quark-Gluon Plasma, Charmonia, Ultra-relativistic Heavy-Ion Collisions}

\begin{document}




\section{Introduction}
\label{sec:intro} 
The production of charmonia in ultra-relativistic heavy-ion collisions (URHICs) is an active area of research since four decades ago. The initially proposed $\jpsi$ suppression signature of quark-gluon plasma (QGP) formation~\cite{Matsui:1986dk} has developed into more comprehensive transport models that account for regeneration mechanisms as dictated by the principle of detailed balance which ensures that the abundances of charmonia approach their pertinent equilibrium limits, see, \eg, Refs.~\cite{Rapp:2008tf,Braun-Munzinger:2009dzl,Liu:2015izf,Andronic:2024oxz} for reviews. 
Abundant charm production at the Large Hadron Collider (LHC), with around 100 charm-anticharm quark pairs in a central Pb-Pb collision at a center-of-mass energy of 5.02\,TeV per nucleon pair, has led to predictions of a substantial amount of regenerated charmonia which have been confirmed by experiment~\cite{ALICE:2016flj,Bai:2020svs,ALICE:2023gco,ALICE:2022jeh,ALICE:2019lga,ALICE:2019nrq}. 
This signature features an approximately constant (or even rising) $\jpsi$ yield with collision centrality in terms of the nuclear modification factor, $\raa$, and a concentration of the regeneration yield at relatively low momenta~\cite{Zhao:2011cv,Song:2011xi,Zhou:2014kka,Ferreiro:2014bia}. In addition, the regenerated charmonia exhibit an appreciable elliptic flow inherited from the recombining charm and anti-charm quarks that have been dragged along with the expanding fireball~\cite{He:2021zej}.
However, significant model uncertainties remain, most notably in the underlying assumptions about the $\jpsi$ dissociation temperature, which controls the onset of regeneration in the cooling fireball, and in the input charm cross section, which determines the equilibrium limit of the charmonia and thus controls the magnitude of the regeneration. 
For example, the Statistical Hadronization Model (SHM) offers a complementary perspective, where all charmonium states are produced via statistical hadronization at a fixed common temperature corresponding the pseudo-critical temperature of the chiral cross-over transition, $T_{\rm pc} \simeq 160$\,MeV~\cite{Andronic:2021erx} (see also Ref.~\cite{Capellino:2023cxe}), while most transport models are based on a hierarchy of dissociation temperatures that is correlated with the charmonium binding energies. 
Both transport and SHM models are quite sensitive to the amount of charm-anti-charm quark pairs in the fireball. Fortunately, the experimental knowledge about the total charm cross section in proton-proton (pp) collisions has much advanced in recent years. With higher precision and an improved assessment of the contribution of charm baryons a noticeable increase in the value of the cross section has emerged~\cite{ALICE:2021dhb}.

Successful measurements of the excited state, $\psip$, in heavy-ion collisions were conducted at the Super Proton Synchrotron (SPS)~\cite{RAMELLO1998261c}. Its strong suppression has been 
explained by both the SHM~\cite{Andronic:2007bi}
and by transport models~\cite{Sorge:1997bg,Spieles:1999kp,Grandchamp:2002wp,Linnyk:2008hp}.
In the latter, the small binding energy of the $\psip$ (about 60\,MeV in vacuum) implies that it has a much smaller dissociation temperature than the $\jpsi$ (with a vacuum binding energy of $\sim$630\,MeV), and thus its in-medium kinetics is operative at later stages in the fireball evolution. In this regard, small collision systems, \ie,  d-Au(0.2\,TeV) collisions at the Relativistic Heavy-Ion Collider (RHIC)~\cite{PHENIX:2013pmn} and p-Pb collisions at the LHC~\cite{ALICE:2014cgk}, turned out to give valuable constraints on the reaction rates of the $\psip$. The expected smaller initial temperatures in these systems did not
cause significant $\jpsi$ suppression beyond CNM effects, while the stronger suppression of the $\psip$ has been interpreted as being due to final-state effects in the more dilute phases of these collisions, relative to Au-Au or Pb-Pb collision systems~\cite{Ferreiro:2014bia}. This allowed for a much improved gauge of the $\psip$ reaction rate~\cite{Du:2015wha,Du:2018wsj}. 
When deployed to heavy-ion collisions, this has led to the notion of a ``sequential regeneration" of $\jpsi$ and $\psip$ mesons~\cite{Du:2015wha}. An initial application to CMS data in Pb-Pb (5.02\,TeV) collisions at the LHC~\cite{CMS:2014vjg} involved a cut on the transverse-momentum of the charmonia of $\pT\ge6.5\,{\rm GeV}/c$, and thus was not directly probing the prevalent regime of the regeneration contributions. This has been improved by recent ALICE data~\cite{ALICE:2022jeh}, which will play a key role in what follows below.

In the present paper, we update our model for quarkonium kinetics in heavy-ion collisions~\cite{Grandchamp:2002wp,Zhao:2010nk} in several respects. Most significantly, we will implement the most recent experimental values for the total charm cross section and refine our treatment of CNM effects (including their $p_T$ dependence); we will also utilize an improved input for the in-medium charmonium binding energies as to ensure an approximately constant $\jpsi$ mass, as was recently done in our calculations for bottomonium transport~\cite{Du:2017qkv}, and re-assess the relevance of inelastic-scattering versus gluo-dissociation mechanisms. In our applications to phenomenology, we will specifically elaborate on our previous predictions for recent ALICE data on $\psip$ production in 5.02\,TeV Pb-Pb collisions at the LHC~\cite{ALICE:2022jeh}, thereby reiterating the role that sequential regeneration plays in interpreting these data.

This paper is organized as follows: In Sec.~\ref{sec:kin}, we briefly recall our calculations of charmonium reaction rates and how they figure in the kinetic-rate equation within a schematic fireball for Pb-Pb collisions at the LHC; we also pinpoint updates specific to this work, \eg, in-medium charmonium binding energies, the total charm cross section with corrections from nuclear shadowing and bottom-decay feeddown in the nuclear modification factors. In Sec.~\ref{sec_time-centrality} we discuss the time dependence of charmonia kinetics in central and peripheral Pb-Pb collisions at the LHC, as well as the centrality dependence of their inclusive yields with comparisons to data. In Sec.~\ref{sec:pt}, we evaluate the $\pT$ dependence of $\jpsi$ and $\psip$ production in Pb-Pb (5.02\,TeV) collisions, based on 
fits to pp spectra. This analysis encompasses $\pT$ spectra, pertinent nuclear modification factors across different centralities, centrality dependent yields within different momentum bins, and the average $\pT$ its square, with comparisons to data as available.
We summarize and conclude in Sec.~\ref{sec:concl}.

\section{Kinetic Approach}
\label{sec:kin} 
In this section, we recall the basic components of our transport approach. We introduce the kinetic-rate equations and their transport parameters in Sec.~\ref{ssec_trans}, give a detailed discussion of the reaction rates in Sec.~\ref{ssec_rates} and of the equilibrium limits in Sec.~\ref{ssec_equil}, and specify the initial conditions and underlying medium evolution in Sec.~\ref{ssec_ini}. 

\subsection{Transport Parameters}
\label{ssec_trans}
Our starting point is a kinetic-rate equation that describes the time evolution of charmonium yields, $N_{\psi}$, according to~\cite{Zhao:2010nk}
\begin{equation}
   \frac{dN_{\psi}(\tau)}{d\tau}=-\Gamma_{\psi}(T(\tau))\left[N_{\psi}(\tau)-N_{\psi}^{\rm eq}(T(\tau))\right] \  ,
\label{eq:rate_eq}
\end{equation}
where $\Gamma_{\psi}$ is the reaction rate and $N_{\psi}^{\rm eq}$ the equilibrium limit of  state $\psi$. In the present work, we include the lowest three states, $\psi= \jpsi, \psip$ and $\chic$, where the latter represents an average over the three $1P$ states $\chi_{c0}$, $\chi_{c1}$ and $\chi_{c2}$, with average mass $m_\chi=3.543$\,GeV and a total spin degeneracy of 9.

For our purposes below, it is useful to note that the time evolution of $N_{\psi}$ from Eq.~\ref{eq:rate_eq} can be formally decomposed into two distinct processes corresponding to a primordial and a regeneration component. The primordial component refers to the initial $\psi$ yields that undergo suppression in the medium. This suppression is directly given by the loss term, 
$-\Gamma_{\psi}(T(\tau))N_{\psi}(\tau)$, reflecting the exponential suppression of the initially produced $\psi$ states. On the other hand, the regeneration component arises from the thermal production of charmonium states within the medium, determined by the approach towards the chemical-equilibrium limit, $N_{\psi}^{\rm eq}(T(\tau))$. 
The regeneration contribution can thus be defined as the solution to the homogeneous rate equation which starts from a vanishing initial condition. It is equivalent to the difference of the full solution minus the suppression contribution. 
\begin{figure}[t]
   \begin{minipage}[b]{1\linewidth}
      \includegraphics[width=6.8cm]{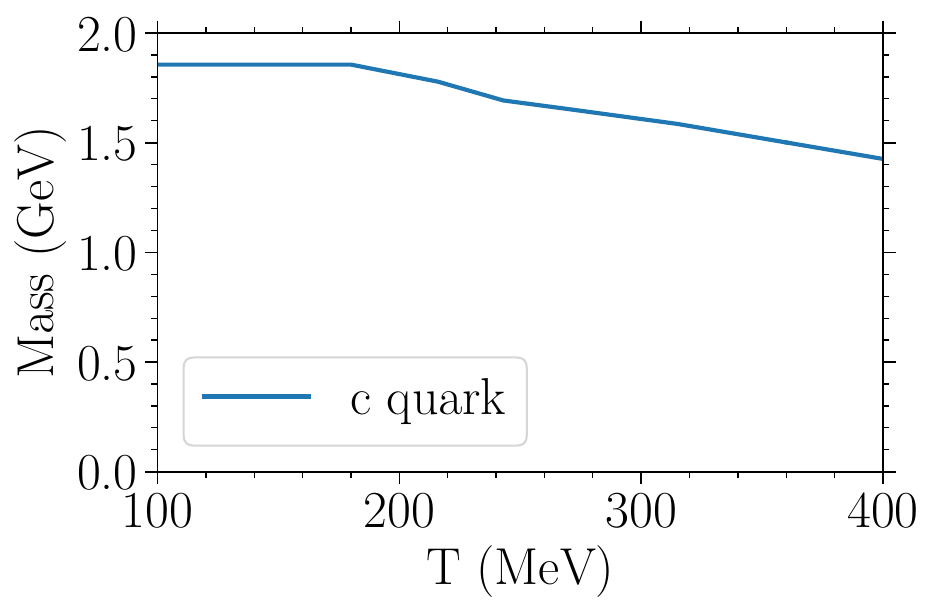}
      \includegraphics[width=6.8cm]{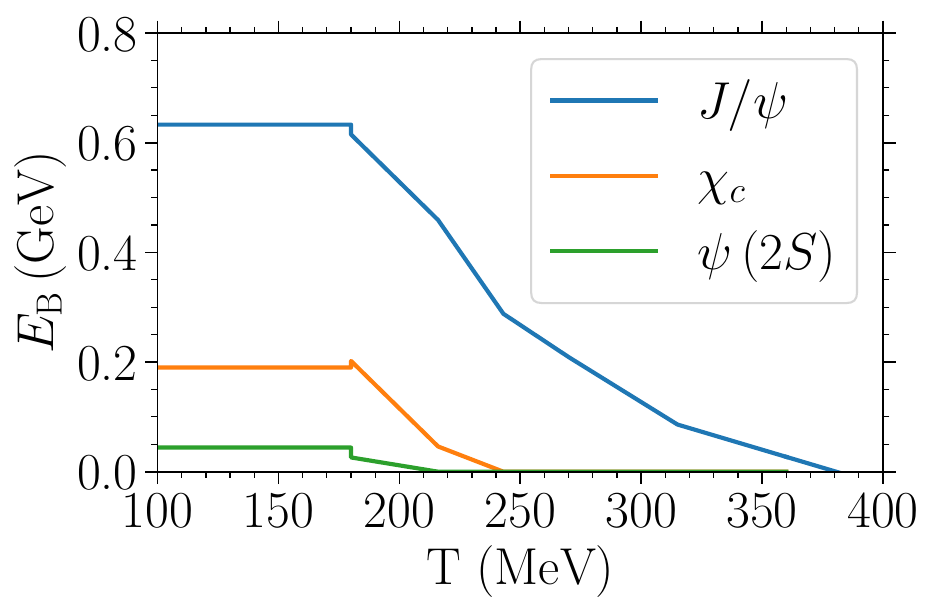}
 \end{minipage}
 \caption{Charm-quark mass (left panel) and charmonium binding energies (right panel) as a function of temperature.}
     \label{fig:mass}
   \end{figure}

The charmonium transport parameters, \ie, reaction rates and equilibrium limits, will be discussed in the following two Secs.~\ref{ssec_rates} and \ref{ssec_equil}.

\subsection{Charmonium Reaction Rates}
\label{ssec_rates}
The reaction rates in the QGP phase are calculated using in-medium charm-quark masses and binding energies guided by the thermodynamic $T$-matrix approach. In an update to previous work~\cite{Zhao:2010nk} where results based on the internal-energy potential from lattice-QCD (lQCD) computations were employed, we have modified the binding energies to ensure that for the given input charm-quark masses, $m_c(T)$, the resulting $\jpsi$ mass is constant with temperature~\cite{Andronic:2024oxz}. This leads to somewhat larger binding energies for temperatures below $\sim$300\,MeV, by up to maximum of $\sim$150\,MeV at $T$$\simeq$220\,MeV, which are, in fact, in better agreement with self-consistent $T$-matrix calculations within the strongly-coupled scenario of Ref.~\cite{Liu:2017qah}. The net effect on observables, is, however quite small, well within other uncertainties in our in input, such as the nuclear shadowing. The inputs are summarized in Figure~\ref{fig:mass}.

\begin{figure}[t]
   \begin{minipage}[b]{1\linewidth}
      \centering
      \includegraphics[width=7.4cm]{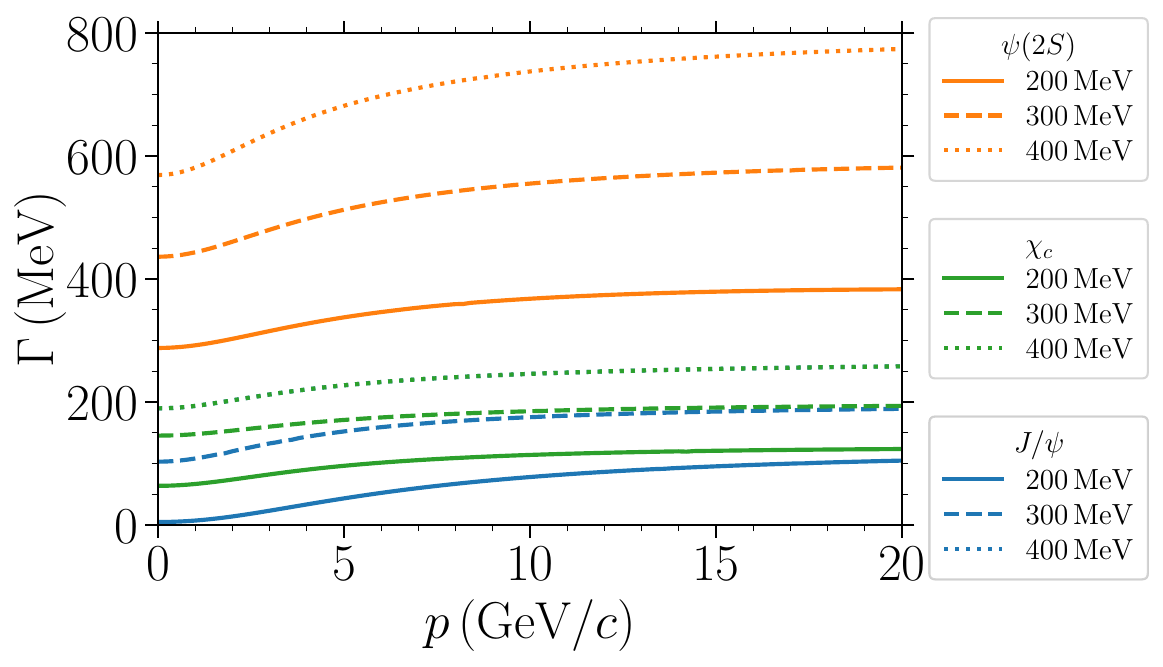}
      \includegraphics[width=6.3cm]{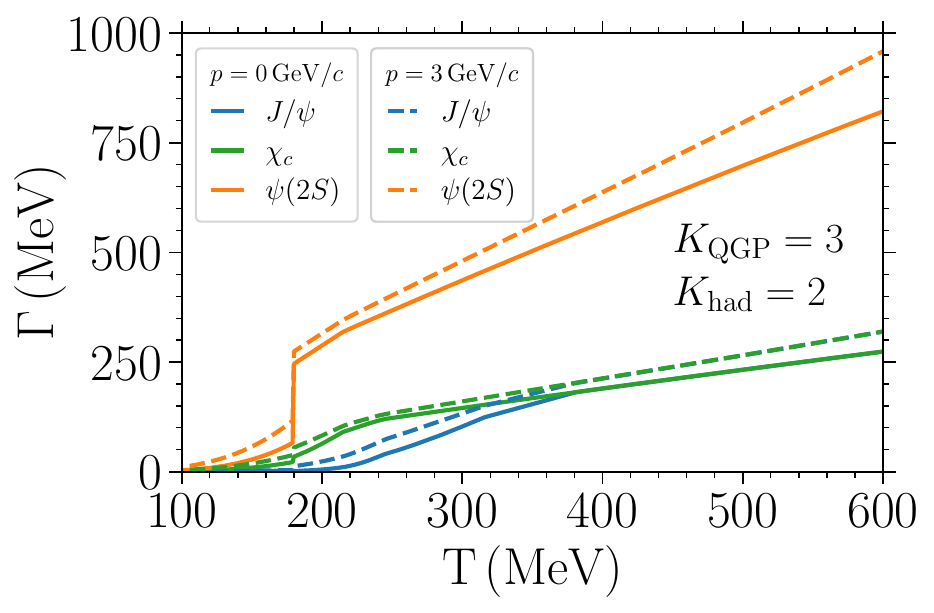}
   \end{minipage}
   \caption{Dissociation rates of charmonia in the medium as a function of momentum at different temperatures (left panel) and temperature at $p$=0 and $3\,{\rm GeV}/c$ (right panel).}
   \label{fig:rate}
\end{figure}

The dominant contribution to the reaction rates arises from inelastic scatterings of thermal partons ($i=q, \bar q, g)$ with charm quarks inside the bound state, \ie, $i+\psi \to c+\bar{c}+i$.
These processes are implemented using perturbative Born diagrams in a quasifree approximation~\cite{Grandchamp:2002wp}, where one of the heavy quarks in the bound state, denoted as $c^*$, is assumed to be half-off-shell, thereby carrying the binding energy. The other quark is treated as a spectator, which essentially amounts to neglecting recoil corrections. The dissociation rate is then determined by a convolution of the inelastic (half off-shell) $2\to2$ cross section (or rather matrix element squared) with a thermal parton distribution function, $f_i$,
\begin{eqnarray}
\Gamma_{\psi}^{\rm qf}(p,T)=2\sum\limits_i\int\frac{\dd^3p_{i}}{(2\pi)^3} f_{i}(\omega_{p_i},T)v^{c^*p} \sigma_{c^*i\rightarrow c i}(s) \ .
\label{eq:quasifree}
\end{eqnarray}
Here, the factor of 2 accounts for the $c$ and $\bar{c}$ quark, with $p$ denoting the momentum of the charmonium. The incoming relative velocity of a $c$ quark and a thermal parton is given by
\begin{equation}
v_{ci}=\frac{\sqrt{\left(p_{c}^{(4)} \cdot p_{i}^{(4)}\right)^{2}-m_{c}^{2} m_{i}^{2}}}{\omega_{c}(p_c) \omega_{i}(p_i)} \ ,
\end{equation}
where $m_i$ denotes the thermal-parton mass and $\omega_{c(i)}$ the on-shell energy of the $c$ quark (thermal parton).
The results of the quasifree rates are summarized in Figure~\ref{fig:rate}; they generically show an increase with 3-momentum, mostly caused by a suppression at low momentum, which is quite sensitive to the binding energies, while a weak increase remains even in the limit of vanishing binding  due to
the increase in final-state phase space. 
For the $\psip$, which is essentially unbound even at low QGP temperatures, it turns out~\cite{Du:2015wha} that coupling the light parton to the bound state using a perturbative scattering diagram is insufficient to describe its suppression observed in d-Au collisions at RHIC~\cite{PHENIX:2013pmn} and p-Pb collisions at the LHC~\cite{ALICE:2014cgk}.
Therefore, the QGP rates for the $\psip$ were augmented by a $K$-factor of 3 to simulate nonperturbative interaction strength~\cite{Du:2015wha}.

\begin{figure}[t]
   \begin{minipage}[b]{1\linewidth}
      \includegraphics[width=6.8cm]{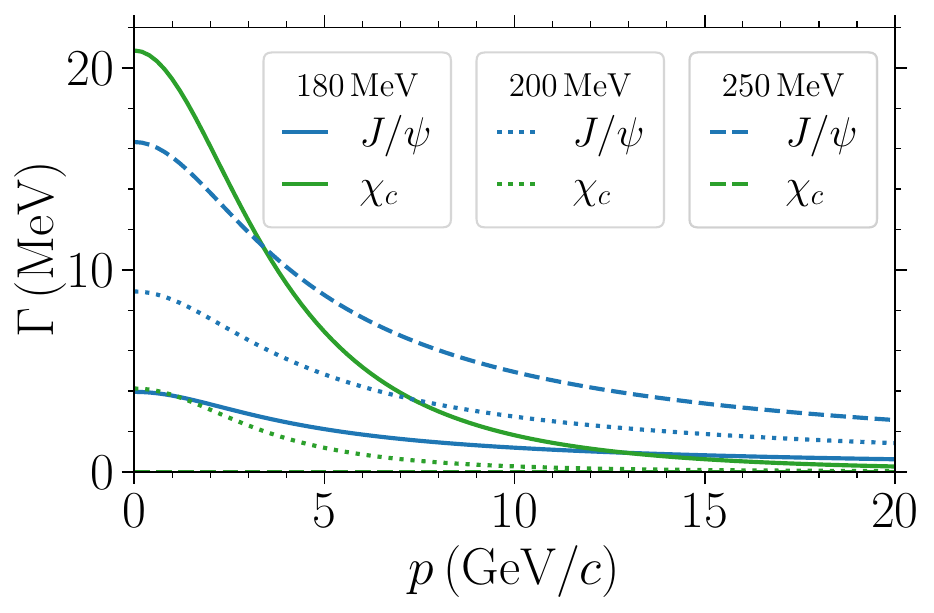}
      \includegraphics[width=6.8cm]{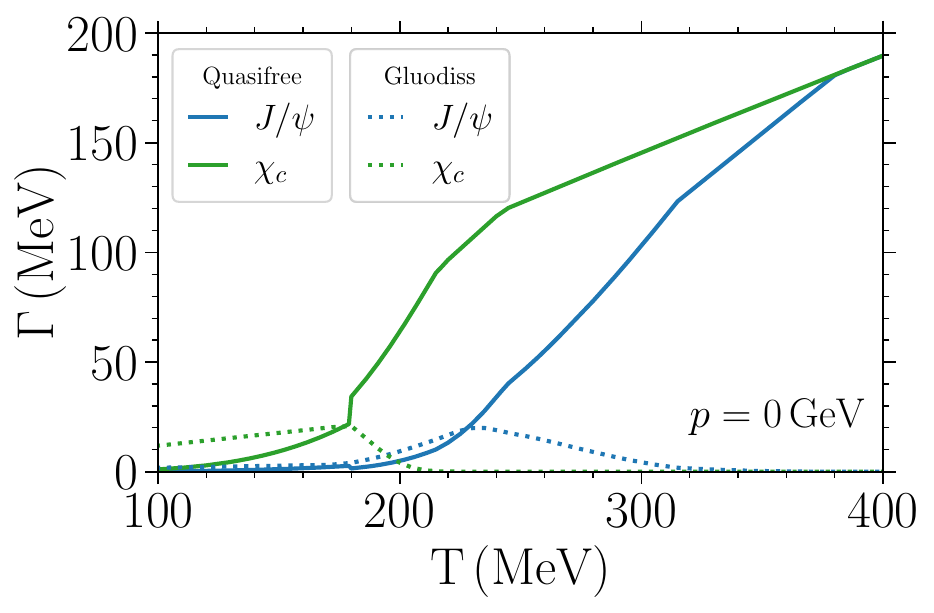}
 \end{minipage}
 \caption{Gluo-dissociation rates of charmonia in the QGP as a function of momentum (left panel) for three different temperatures and as a function of temperature at $p=0$ (right panel) where they are also compared to the quasifree rates.}
     \label{fig:rate_gd}
   \end{figure}

We also revisit the relevance of gluo-dissociation processes, $g+\psi\to c+\bar{c}$,  whose inelastic rate can be written as~\cite{Peskin:1979va,Bhanot:1979vb},
\begin{equation}
   \Gamma_{\psi}^{\rm gd}(p,T)=\int\frac{\dd^3p_{g}}{(2\pi)^3}d_g f_{g}(\omega_{p_g},T)v_{\rm rel} \sigma_{\psi g\rightarrow c \bar{c}}(s) \ .
   \end{equation}
Here, we adopt a slightly different (and, as we believe, more consistent) implementation compared to our previous studies~\cite{Zhao:2010nk}, by following our recent work on bottomonia~\cite{Du:2017qkv}. 
Rather than expressing the cross section entirely in terms of its binding energy (as originally derived for a Coulombic bound state), we write it as
\begin{equation}
\sigma_{\psi g \rightarrow c\bar{c}} = \frac{2\alpha_s}{3 m_c E_B} g_{\psi}(x)
\end{equation}
where the dependence on the strong coupling constant, $\alpha_s$, signifies the perturbative coupling to the timelike gluons from the surrounding heat bath (assumed to be made of massive partonic quasiparticles), while the remaining dependencies on $E_B$ and charm-quark mass characterize the in-medium bound-state properties. Thus, these are implemented on the same footing as the quasifree process. The functions $g_{\psi}(x)$ then take the following form for the three different charmonia we account for:
\begin{equation}
g_{\psi}(x) = \left\{
\begin{array}{ll}
\frac{2}{3\pi}\left(\frac{32}{3}\right)^2 \frac{(x-1)^3}{x^5} & \text{for } \psi=J/\psi \\
\frac{2}{3\pi}\left(\frac{32}{3}\right)^2 \frac{16(x-1)^3}{(x-3)^2 x^7} & \text{for } \psi=\psi(2S) \\
\frac{2}{3\pi}\left(\frac{32}{3}\right)^2 \frac{4(x-1)^2 (9x^2 - 20x +12)}{x^7} & \text{for } \psi=\chi_{c} \ , 
\end{array}
\right.
\end{equation}
where $x = k_0 / E_B$, and $k_0 = \frac{s - m_{\psi}^2 - m_{g}^2}{2m_{\psi}}$ is the incident gluon energy in the rest system of $\psi$. The center-of-mass energy squared,
\begin{equation}
s = (p^{(4)} + p_{g}^{(4)})^2 = m_{\psi}^2 + m_{g}^2 + 2E_{\psi}\omega_{g} - 2\vec{p} \cdot \vec{p}_{g} \ , 
\end{equation}
is obtained from the incoming charmonium and gluon 4-momenta in the thermal system,
\begin{equation}
p^{(4)} = (E_{\psi}, \vec{p}) \quad \text{and} \quad p_{g}^{(4)} = (\omega_{g}, \vec{p}_{g}) \ , 
\end{equation}
respectively.
The gluo-dissociation rates of the $\jpsi$ and $\chi_c$, shown in Figure~\ref{fig:rate_gd}, are negligible compared to the quasifree
rates at temperatures $T\gsim250$\,MeV and $T\gsim190$\,MeV, respectively (even more so at finite 3-momentum). Once the gluo-dissociation rates become comparable or larger than the quasifree rates at lower temperatures, both rates are numerically small, implying that their impact on charmonium transport will be small (this will be quantified below). The gluo-dissociation rates for the $\psip$ (not shown) are negligibly small, as expected from the small binding energy of the $\psip$.

For the rates in hadronic matter, we follow previous developments~\cite{Grandchamp:2002wp,Du:2015wha} where effective SU(4) Lagrangian calculations of meson exchange interactions in pion- and $\rho$-meson induced dissociation~\cite{Lin:1999ad,Haglin:2000ar} were extended to a large set of non-/strange resonances based on phase space considerations. For the $\jpsi$, the resulting rates remain quite small; however, they are significant for the $\psip$, although still significantly smaller than in the QGP (with $K$ factor), as shown in the right panel of Figure~\ref{fig:rate}.
%

\subsection{Charmonium equilibrium limits}
\label{ssec_equil}
Detailed balance enforces the yields of the charmonium states, $N_{\psi}$, to approach their respective equilibrium limits, $N_{\psi}^{\rm eq}$. We evaluate these equilibrium limits through the standard thermal-density expression,
\begin{equation}
   N_{\psi}^{\rm eq}(T) =  V_{\rm FB} d_{\psi} \gamma_c^2\int \frac{\dd^3p}{(2\pi)^3} \exp(-E_{\psi}/T)
   =V_{\rm FB}\frac{d_{\psi}}{2 \pi^2}\gamma_c^2  T m_{\psi}^2  K_2\left(\frac{m_{\psi}}{T}\right)\  ,
   \label{eq:Neq}
\end{equation}
where $d_{\psi}$ is the spin degeneracy factor, $E_{\psi}=\sqrt{p^2+m_{\psi}^2}$ the charmonium energy, $V_{\rm FB}$ denotes the (time-dependent) fireball volume, and $K_2$ is the modified Bessel function of the second kind.
The charm-quark fugacity, $\gamma_c$, is computed as in previous work of our approach~\cite{Zhao:2010nk,Du:2017qkv},
assuming conservation of $c\bar{c}$ pairs during the expansion of the fireball,
\begin{equation}
\label{eq:Ncc}
N_{c\bar c}=\frac{1}{2}\gamma_{c} n_{\rm{op}}V_{\rm{FB}}\frac{I_1(\gamma_{c} n_{\rm{op}}V_{\rm{FB}})}
{I_0(\gamma_{c} n_{\rm {op}}V_{\rm{FB}})} + \gamma_{c}^2 n_{\rm{hid}} V_{\rm{FB}}  \ , 
\end{equation}
where $n_{\rm{op}}$ and $n_{\rm{hid}}$ are the densities of open- and hidden-charm states at a given temperature $T$ (charm quarks in the QGP phase or charm hadrons in hadronic matter, with the contribution from charmonia being rather negligible);
$I_0$ and $I_1$ are the modified Bessel functions of the first kind.
The fugacities are matched to the number 
of charm-anticharm quark pairs, $N_{c\bar{c}}$, produced in primordial nucleon-nucleon collisions (accounting for shadowing corrections) and evaluated in the following section.

To account for the non-thermal distributions of charm quarks in the expanding fireballs of URHICs, which tend to suppress the regeneration contribution~\cite{Du:2022uvj}, we adjust the equilibrium limit with a relaxation time factor~\cite{Grandchamp:2002wp}, ${\cal R} = 1-\exp\left(-\int\limits_0^\tau d\tau'/\tau_c\right)$, where the charm-quark thermalization time, $\tau_c$, is taken as $4.5 {\rm fm}/c$, representing an approximate average over 3-momentum and temperature (see, \eg, Fig.~3.3 in Ref.~\cite{He:2022ywp}).

\subsection{Initial Conditions and Medium Evolution}
\label{ssec_ini}
%
\begin{table}[!b]
\centering
\caption{Charm/onium cross sections and $\psip$ over$\jpsi$ ratio  for pp collisions at 5.02 TeV . 
 }
\label{tab:cross_sections}
\begin{tabular}{lcc}
\hline
\textbf{cross section} & \textbf{Mid-rapidity} & \textbf{Forward rapidity} \\
\hline
$\dd\sigma_{c\bar{c}}/\dd y$ (mb) & $1.165 \pm 0.133$~\cite{ALICE:2021dhb,ALICE:2023sgl}& $0.72 \pm 0.07$\\ 
$\dd\sigma_{\jpsi}/\dd y$ ($\mu$b) & $5.64$~\cite{ALICE:2019pid} & $3.93$~\cite{ALICE:2021qlw} \\
\hline
\multicolumn{3}{l}{\textbf{Direct-production cross section ratio in pp collisions}} \\
\multicolumn{2}{l}{\(N_{\psip}^{pp}/N_{\jpsi}^{pp}\)} & \(0.147\)~\cite{ALICE:2021qlw} \\
\hline
\end{tabular}
\end{table}

\begin{figure}[t]
   \centering
   \includegraphics[width=6.8cm]{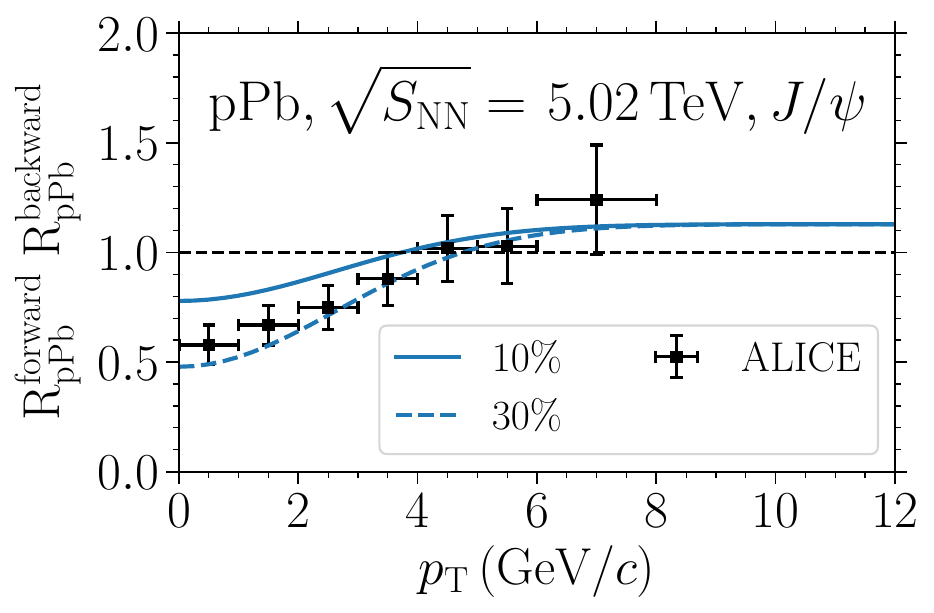}
   \includegraphics[width=6.8cm]{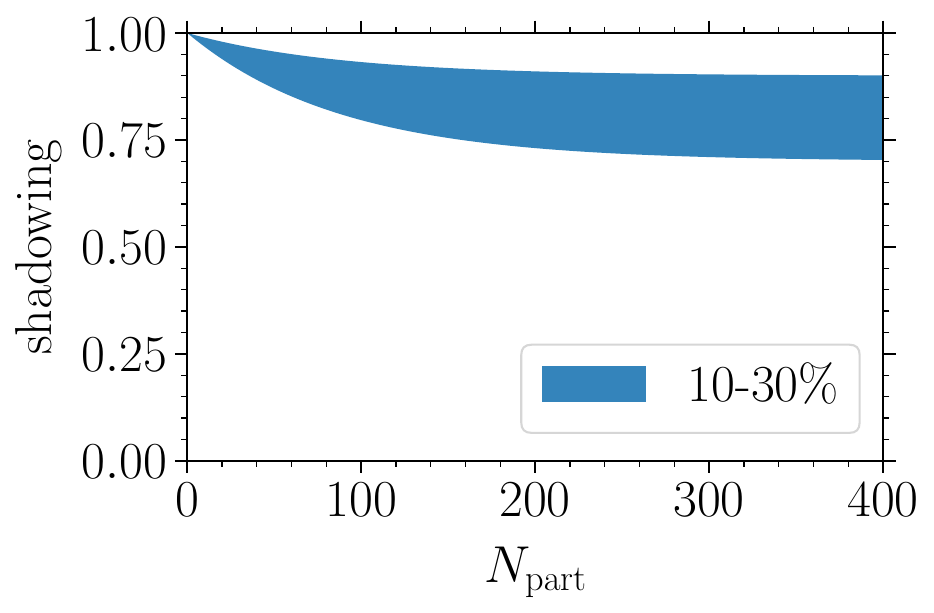}
 \caption{Our parameterizations of the suppression of the $c\bar c$ production cross section due to nuclear shadowing of the parton distribution functions, as a function of $\pT$ (left panel) in p-Pb collisions in terms of the product at forward and backward rapidities and compared to ALICE data~\cite{ALICE:2015sru}, and versus $\npart$ in Pb-Pb collisions (right panel).}
   \label{fig:shadowing}
\end{figure}
The cross section for $c\bar c$ pair production in pp collisions is a key input for the equilibrium limits and thus controls the amount of regeneration. It is usually applied for a specific rapidity interval as 
$N_{c\bar c} = \dd \sigma_{c\bar{c}}/\dd y \Ncoll$, 
where $\Ncoll$ denotes the number of binary nucleon-nucleon collisions at a given collision energy and centrality (estimated from the optical Glauber model~\cite{Miller:2007ri}; for the partonic production processes under consideration here, which are mostly driven by gluon fusion, no distinction is made for the proton-neutron and neutron-neutron collisions). Similarly, we obtain the initial number of charmonium states, which are required to determine the initial condition for the rate equation. 
In Table~\ref{tab:cross_sections} we summarize the charm/onium cross sections for pp collisions at 5.02\,TeV and the ratio of $\psip$ to $\jpsi$ (we note that our values for the charm cross section, taken from Ref.~\cite{ALICE:2021dhb}, are well in line with the most recent assessment in Ref.~\cite{ALICE:2023sgl}). 
The open-charm cross section at forward rapidity has been deduced from its counterpart at midrapidity through the rapidity dependence outlined in Ref.~\cite{Bierlich:2023ewv}. This extrapolation yields a value $0.72 \pm 0.07$, where we incorporate an uncertainty of $\pm10\%$.


A modification of the charm(onium) cross section due to cold-nuclear-matter effects, commonly referred to as nuclear shadowing, is estimated using ALICE data~\cite{ALICE:2015sru}
on $\jpsi$ production in p-Pb collisions at forward and backward rapidities, $2.5 <\left|y\right| < 4$. We fit the product of the measured forward and backward nuclear modification factors, $R_{\rm pPb}$, which can be interpreted as the net effect of shadowing in a Pb-Pb collision, as shown in the left panel of Figure~\ref{fig:shadowing}. Our fit is not inconsistent with recent nuclear parton distribution functions, see, \eg, Ref.~\cite{Kusina:2017gkz}. In addition, we use the same parameterization at mid-rapidity, which is compatible with recent ALICE data as well~\cite{ALICE:2021lmn}. 
In earlier applications of our transport approach to p-Pb collisions at the LHC~\cite{Du:2018wsj}, where a short-lived QGP is predicted to be formed, it was found that about 10-20\% of the inclusive $\jpsi$ suppression is from hot-matter effects (primarily from feeddown of suppressed excited states), which must be ``corrected" for when assessing the shadowing effect. Therefore, we adopt a baseline of 10-30\% suppression of the integrated yield stemming from shadowing, with a $\pT$ dependence that reproduces the forward-backward $R_{\rm pPb}$ product. The $\npart$-dependence of the shadowing for the $c\bar{c}$ cross section is displayed in the right panel of Figure~\ref{fig:shadowing}.

In our comparisons to experimental results for Pb-Pb collisions discussed below, we incorporate the experimental uncertainties of the charm-anticharm cross section $\sigma_{c\bar{c}}$, denoted by  $\Delta \sigma_{c\bar{c}}$, along with the shadowing factor $S$ and its uncertainty, $\Delta S$. The uncertainty of the pp cross section and its shadowing is combined into an effective cross section, $\tilde{\sigma}_{c\bar{c}}$, as follows:
\begin{equation}
\tilde{\sigma}_{c\bar{c}}=\left(S\pm \Delta S\right)\left(\sigma_{c\bar{c}}\pm\Delta \sigma_{c\bar{c}}\right)\simeq S \sigma_{c\bar{c}} \pm \Delta S \sigma_{c\bar{c}}\pm S\Delta \sigma_{c\bar{c}}= \left(S\pm \Delta \tilde{S}\right) \sigma_{c\bar{c}} \ ,
\end{equation}
where $\Delta \tilde{S}$ is determined by the expression
\begin{equation}
\Delta \tilde{S} = S\sqrt{\left(\frac{\Delta S}{S}\right)^2 + \left(\frac{\Delta \sigma_{c\bar{c}}}{\sigma_{c\bar{c}}}\right)^2} \ .
\label{eq:uncertainty}
\end{equation}
The resulting uncertainty bands will be displayed in the figures in the subsequent sections, to provide a visual quantification of the total uncertainty involved.
\begin{figure}[t]
   \centering
   \includegraphics[width=9.5cm]{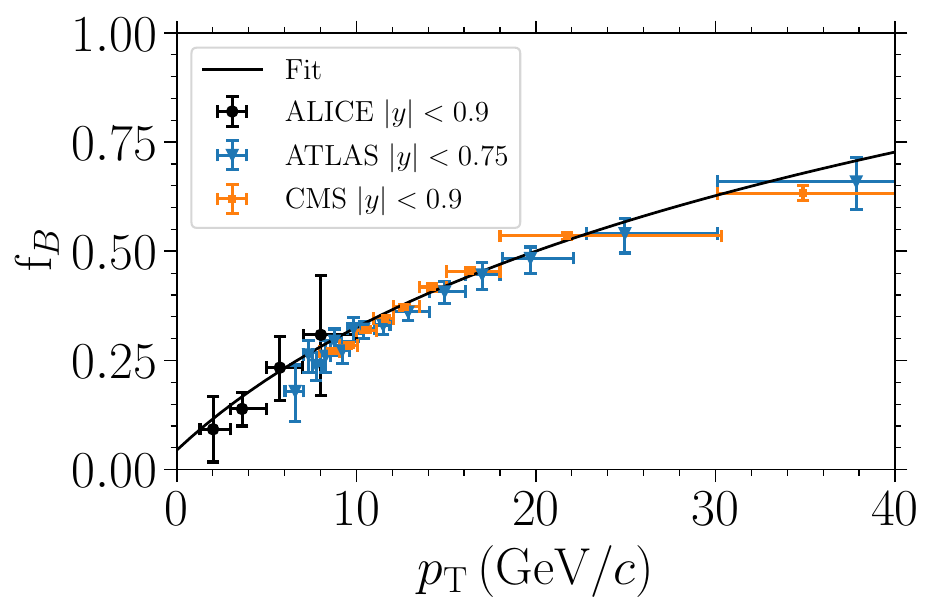}
   \caption{Fraction of bottom-hadron decay feeddown to inclusive $\jpsi$ production as measured by ALICE, ATLAS and CMS in pp collisions at the LHC~\cite{Andronic:2015wma}; our functional fit is shown by the solid line.
   }
   \label{fig:bf}
\end{figure}

Data for inclusive $\jpsi$ production also include feeddown contributions from final-state decays. For the ``prompt" feeddown from electromagnetic and strong decays of excited states, we account for 8\% and 25\% from $\psip$ and $\chic$ mesons, respectively~\cite{Andronic:2015wma}.
Additionally, there is also a ``non-prompt" fraction from bottom decay feeddown (unless explicitly removed experimentally).
We estimate this fraction from available data~\cite{Andronic:2015wma} in pp collisions at the LHC, cf.~Figure~\ref{fig:bf}. It amounts to around 5\% at $\pT$=0 and increases to approximately 50\% at $\pT=20\,{\rm GeV}/c$.
We fit the data using the empirical parameterization
\begin{equation}
   f_{\rm B}(\pT)=0.47 log(0.09\pT+1.1) \ .
   \label{eq:bfd}
\end{equation}

\begin{figure}[t]
   \centering
   \includegraphics[width=9.5cm]{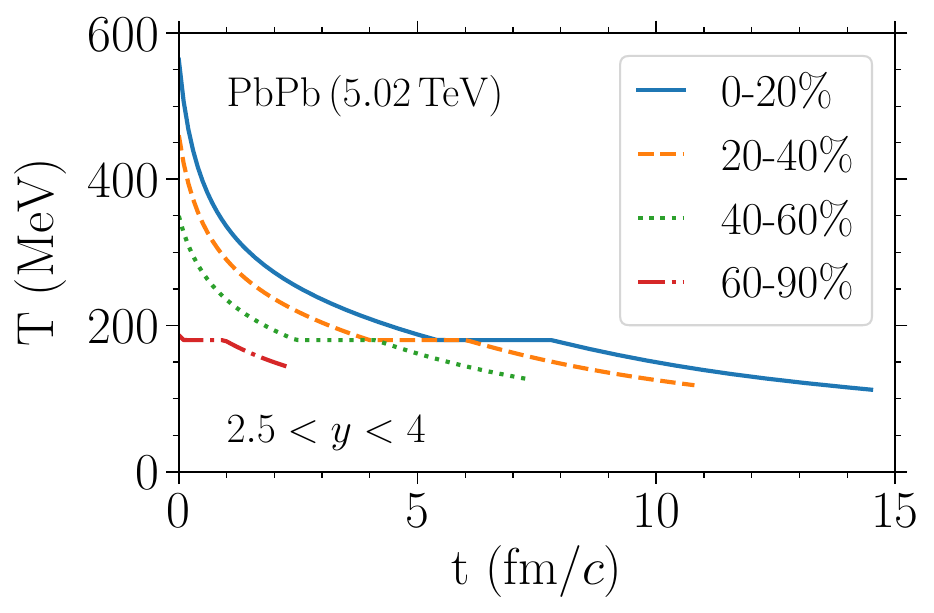}
 \caption{The temperature evolution of the fireball model at forward rapidity in 5.02\,TeV Pb-Pb collisions at different centralities.}
\label{fig:temp}
\end{figure}

The implementation of the rate equation into URHICs requires the space-time evolution of the volume and temperature of the expanding medium. Toward this end, we employ a rather simple fireball model for a cylindrical, isotropic, and isentropic evolution, as elaborated in previous studies~\cite{Grandchamp:2002wp,Zhao:2010nk,Zhao:2011cv} for SPS, RHIC and LHC energies. The volume expansion, reminiscent of hydrodynamic models, essentially corresponds to a time-dependent blastwave model with a collective flow at thermal freezeout ($T_{\rm fo}\simeq$~100\,MeV in central Pb-Pb collisions) that reproduces measured $p_T$ spectra of light hadrons. The total entropy in the fireball, calculated from the observed multiplicity of charged particles using a hadron resonance gas model, is assumed to remain constant throughout the adiabatic expansion. By monitoring the entropy density, $s(\tau)=S_{\rm tot}/V_{\rm FB}$, at each time $\tau$ we can infer the fireball temperature once we specify the
equation of state (EoS) for the medium, \ie,  $s(T)$.
For the QGP phase, we adopt an ideal gas of massive quarks and gluons, while the hadronic phase is represented by a non-interacting gas of resonant states, including mesons and anti-/baryons with masses of up to 2\,GeV. The critical temperature is set to $T_c$=180\,MeV (we have checked that employing a more realistic EoS based on lattice-QCD data, with a continuous transition into a hadron resonance gas, has negligible impact on quarkonium kinetics~\cite{Du:2017qkv}). While the assumption of an isotropic volume, $V_{\rm FB}$, without spatial temperature gradients is rather schematic, it enables 
a straightforward calculation of the charm-quark fugacity factor, $\gamma_c$ in Eq.~(\ref{eq:Ncc}). The latter governs the equilibrium limit, Eq.~(\ref{eq:Neq}), of the different charmonium states and is therefore key in obtaining a reliable gain term for the regeneration contribution. The extension to individual cells with different temperatures, as figuring, \eg, in a hydrodynamic evolution, renders this more challenging, but will be addressed in future work.  
Figure~\ref{fig:temp} illustrates the resultant temperature evolution as a function of proper time across various centralities at LHC energies int he forward rapidity region. At central rapidities, where the observed charged-particle multiplicities are about 20\% larger for the same centrality class, the initial temperatures increase by about 6\%.  

Right after the near-instantaneous formation of a $c\bar{c}$ pair which would develop into a $\psi$ in a pp collision, we incorporate initial formation time effects to approximate the (quantum) expansion dynamics of evolving into a fully formed bound state. Contrary to expectations derived from a classical cross section perspective where the transverse size (\ie, cross sectional area) would grow quadratically with time, we utilize a scaling that is linear with time~\cite{Farrar:1988me} and scale down the reaction rates by a factor of $\tau/\tau_{\rm form}$ for $\tau \leq \tau_{\rm form}$. The quantum formation times are estimated based on energy uncertainties associated with the splitting in binding energies, \ie,  $\tau_{\mathrm{form}}(\jpsi,\psip,\chic)=1,2,2\,\text{fm}/c$, respectively.

Concerning regeneration processes, we assume their onset once the cooling medium has reached the pertinent dissociation temperature, \ie, the point where the binding energy vanishes, $T_{\rm diss} \simeq 180, 240, 360$~MeV for $\psip$, $\chic$ and $\jpsi$, respectively. 
Quantum mechanical uncertainty suggests that bound states are distinctly defined only when their binding energies are on the order of or greater than their respective width. 
However, even for smaller (or vanishing) $E_B$ values, resonance-like correlations can persist, potentially facilitating the population of the relevant quantum states.
A more elaborate treatment of this regime, also referred to as the quantum-Brownian motion regime, as well as of the formation time effects referred to above, necessitates a quantum-transport approach.

\section{Time and Centrality Dependence of Charmonium Yields at the LHC}
\label{sec_time-centrality}
Primordial heavy-quark(onium) production in URHICs is expected to scale with the number of binary nucleon-nucleon collisions upon initial impact, $\Ncoll$. To quantify medium effects as a deviation from this expectation, it is a common practice to analyze (the modification to) the quarkonium production yields in terms of the nuclear-modification factor, $\raa$, defined as
\begin{equation}
\raa^\psi(\npart) = \frac{N^{AA}_\psi(\npart)}{N_\psi^{pp} \Ncoll(\npart)}  \ ,  
\end{equation}
where $N_\psi^{pp}$ denotes the inclusive $\psi$ yield in pp collisions at the same collision energy. The number of nucleon participants, $\npart$, is estimated from a Glauber model for a given impact parameter, $b$~\cite{Miller:2007ri}, and serves as a measure of the centrality of the nuclear collision.
Unless otherwise stated, the denominator of the $\raa$ will include both prompt and non-prompt feeddown contributions and utilize the central values for the input cross sections at given rapidity, and the reaction rates are for quasifree dissociation.

In the remainder of this section, we focus on $\pT$-integrated yields using a 3-momentum averaged reaction rate obtained by solving the rate equation, Eq.~\ref{eq:rate_eq}; we first discuss the time evolution of direct $\jpsi$ and $\psip$ production for two specific centralities in Sec.~\ref{ssec_time-evo}, and then turn to the centrality dependence of inclusive yields in comparison to experiment in Sec.~\ref{ssec_centrality}, including the recently measured $\psip/\jpsi$ ratio. 


\subsection{Time evolution of charmonium yields}
\label{ssec_time-evo}
%
\begin{figure}[!t]
   \centering
   \includegraphics[width=0.98\textwidth]{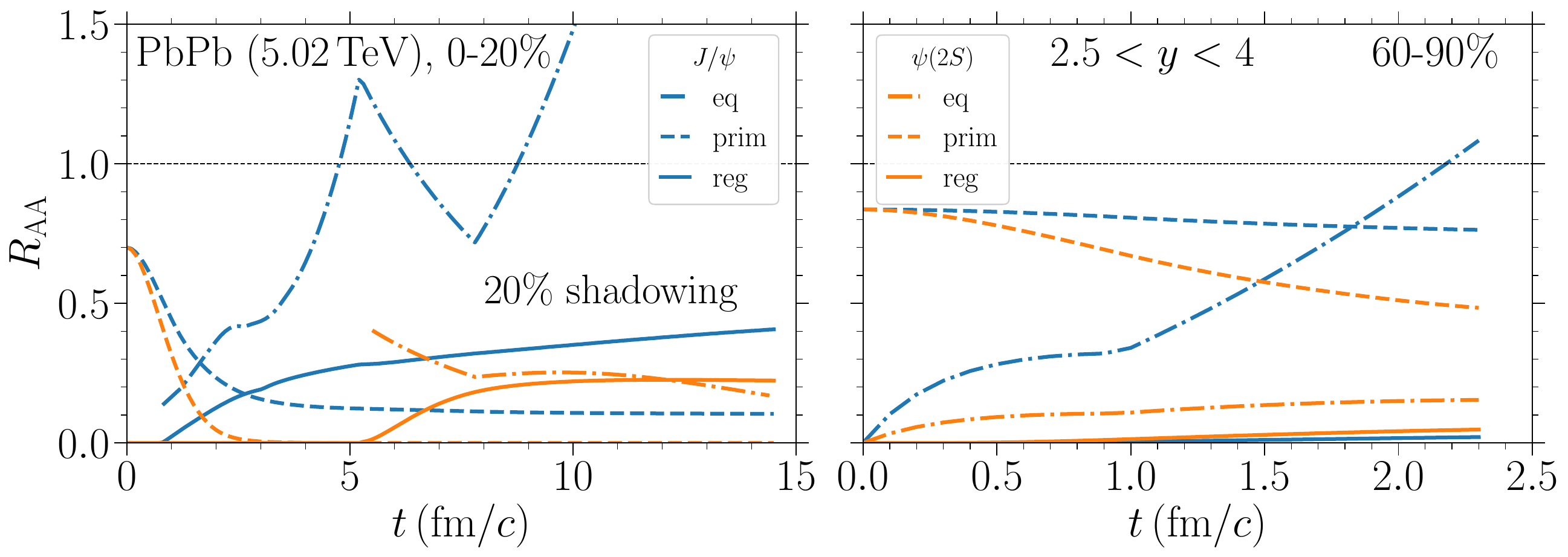}
 \caption{Time evolution of charmonium kinetics in central (left panel) and peripheral (right panel) Pb-Pb collisions at the LHC at forward rapidity. Blue and orange lines represent the direct $\jpsi$ and $\psip$ yields, respectively, where the solid and dashed lines correspond to the suppressed primordial and regenerated contributions, respectively, while the dashed-dotted lines are the pertinent equilibrium limits (including the thermal relaxation time correction). The calculations are carried out a charm cross section of $\dd \sigma_{c\bar{c}}/\dd y =0.72\,$mb (the central value at forward rapidity, recall Table~\ref{tab:cross_sections}) including an up to 20\% shadowing in central collisions (which is down to about 4\% at 60-90\% centrality).}
     \label{fig:time_evo}
\end{figure}
To illustrate the time evolution of the nuclear-modification factors for $\jpsi$ and $\psip$, we focus on 0-20\% and 60\%-90\% central Pb-Pb (5.02\,TeV) collisions at forward rapidity, cf.~Figure~\ref{fig:time_evo}. We include contributions from both primordial and regenerated yields, alongside their equilibrium limits, where the numerator only includes direct production (\ie, excluding prompt of weak-decay feeddown).
In the initial phases of the QGP evolution in central collisions, both primordial charmonium states undergo strong suppression, with the $\psip$ yield essentially being wiped out. 
In contrast, peripheral collisions show significantly less suppression, especially for the $\jpsi$, due to a lower fireball temperature and shorter lifetime. Nevertheless, a marked suppression of the $\psip$ is still operative as its reaction rates are still appreciable at the critical temperature as well as in the hadronic phase.

In central collisions, the regeneration of the $\jpsi$ starts well within the QGP phase, but never really reaches the equilibrium limit, especially in the later stages where the reaction rates are too small (although the large equilibrium limit still produces a small contribution from regeneration). On the other hand, the $\psip$ regenerates significantly later, because of its smaller dissociation temperature and the associated larger reaction rates, reaching (and sustaining) its equilibrium limit towards the end of the mixed phase (and into the hadronic phase). 
As a consequence of the ``sequential regeneration" of $\jpsi$ and $\psip$, their final ratio surpasses the pertinent equilibrium limit at any given temperature. 
In peripheral collisions, both charmonium states commence regeneration concurrently. However, the $\psip$ is subject to significantly higher rates compared to the $\jpsi$ which leads to a larger regeneration-$\raa$, although quantitatively still small owing to the relatively low equilibrium limit. 

\begin{figure}[!t]
   \centering
   \includegraphics[width=13cm]{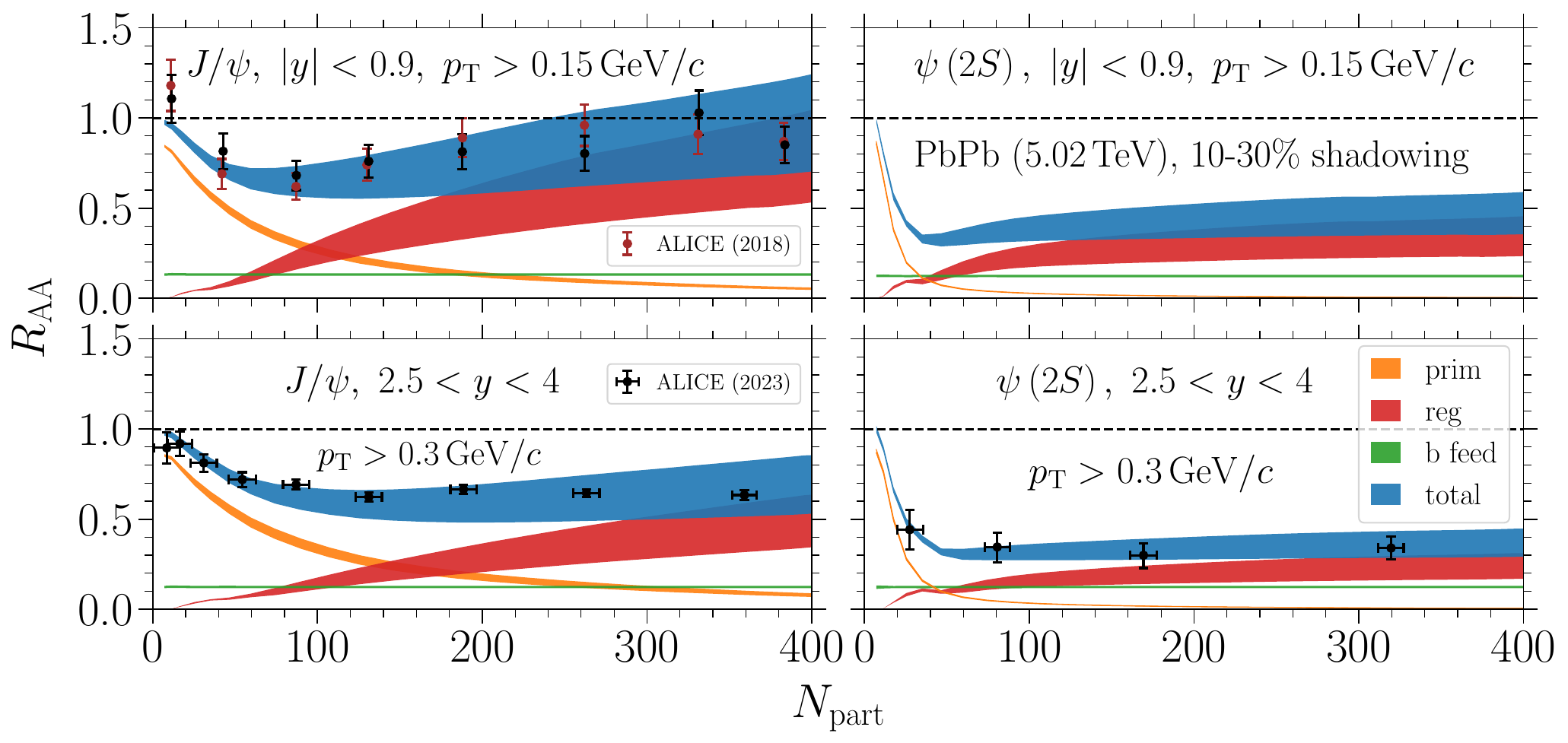}
   \caption{
   Centrality dependence of charmonium $\raa$'s in Pb-Pb(5.02\,TeV) collisions at the LHC. The upper (lower) panels are for mid- (forward) rapidities and the left (right) panels are for $\jpsi$ ($\psip$). 
   The bands for the primordial (orange), regenerated (red), and total (blue) components include uncertainties from the initial charm cross section and the shadowing effect (added in quadrature). The experimental cuts in pair $\pT$ serve to suppress the contribution from coherent photoproduction and are also applied to our calculations based on the $\pT$ spectra computed in Sec.~\ref{sec:pt}. The calculations are compared to ALICE $\jpsi$ data from 2018 (brown) and 2023 (black)~\cite{ALICE:2016flj,Bai:2020svs,ALICE:2023gco}, and $\psip$ data~\cite{ALICE:2022jeh} data.
   }
   \label{fig:npart}
\end{figure}
%
\subsection{Centrality dependence}
\label{ssec_centrality}
%
The centrality dependence of $\jpsi$ and $\psip$ yields is obtained by evaluating Eq.~\ref{eq:rate_eq} with initial conditions determined for a given (average) participant number, complemented with prompt and non-prompt feeddown contributions. The results for inclusive $\jpsi$ and $\psip$ production in Pb-Pb (5.02\,TeV) collisions, as a function of $\npart$ at mid- and forward rapidity, are summarized
in Figure~\ref{fig:npart}.
The $\jpsi$ yield exhibits the well-established behavior at LHC energies: a strong initial suppression that sets in rather gradually with centrality and is taken over by regeneration contributions at participant numbers of around 100-150. At mid-rapidity, both suppression and regeneration are slightly stronger than at forward rapidity, due to a hotter medium and a larger charm cross section, respectively. The interplay of these mechanisms produces a fairly flat centrality dependence for the total $\raa$, with a mild rise at mid-rapidity, again due to the larger charm production. Overall, the ALICE $\jpsi$ data are reasonably well described~\cite{ALICE:2016flj,ALICE:2022jeh,ALICE:2023gco}, with a preference for shadowing on the weaker side of our central values. For the $\psip$, the right panels in Figure~\ref{fig:npart} show our predictions based on Refs.~\cite{Zhao:2011cv,Du:2015wha} with the updated inputs as discussed in the previous section. Compared to the $\jpsi$, a much steeper suppression of the initial production is found due to the larger rates and smaller dissociation temperatures, leading to a near-complete suppression for $\npart\gsim$~100. Detailed balance causes a regeneration contribution that takes over from the primordial yield at $\npart$ as low as $\sim$40. The resulting inclusive $\raa$ is also quite flat, but leveling off at a substantially smaller value than for the $\jpsi$, chiefly due to the smaller equilibrium limit caused by its larger mass.

\begin{figure}[!t]
   \centering
   \includegraphics[width=9.5cm]{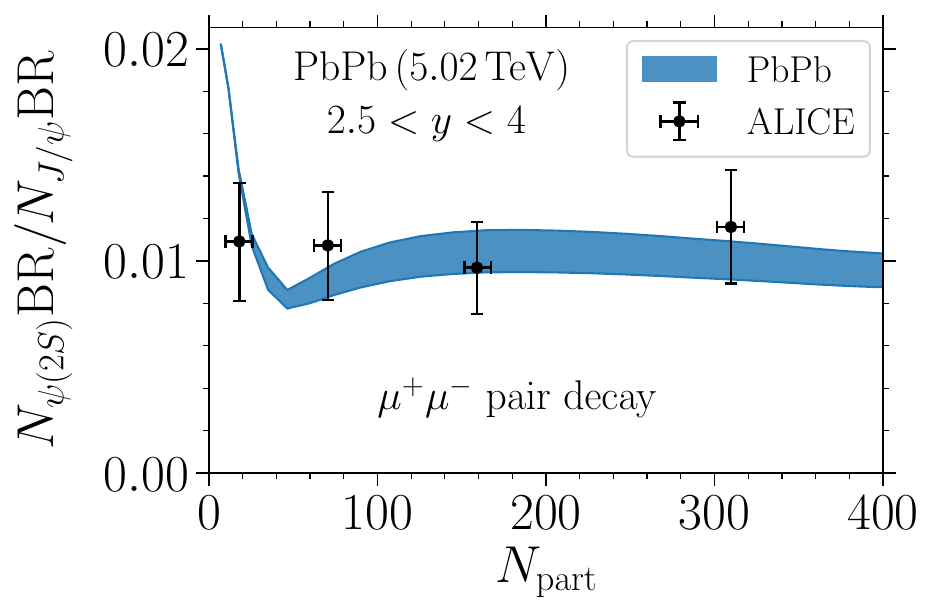}
     \caption{The ratio of $\psip$ over $\jpsi$ as a function of $\npart$ in Pb-Pb(5.02\,TeV) collisions at forward rapidity, compared to ALICE data~\cite{ALICE:2022jeh}. The bands indicate the uncertainty of the $\psip$ dissociation temperature around the mixed phase, $\Td=179-180$\,MeV.}
     \label{fig:jpsi_psip}
\end{figure}
We also evaluate the ratio of $\psip$ to $\jpsi$, which mitigates some of the uncertainties in the individual yields, such as the input charm cross section. To illustrate the uncertainty caused by the assumption of the $\psip$ dissociation temperature (below which regeneration sets in), we vary it in the range of 179-180\,MeV (at the beginning of the mixed phase, which is our default) as a lower and upper limit, respectively.
Using the pertinent branching fractions for dilepton decays of $\text{BR}(\psip \rightarrow \mu^{+} \mu^{-}) = 8 \times 10^{-3}$ and
$\text{BR}(\jpsi \rightarrow \mu^{+} \mu^{-}) = 5.961 \times 10^{-2}$~\cite{ParticleDataGroup:2020ssz}, the predicted 
$\npart$ dependence is shown in Figure~\ref{fig:jpsi_psip}.
After an initial rather sharp drop in peripheral collisions (driven by the strong suppression of primordial $\psip$'s)
the ratio essentially saturates and turns out to be consistent with the experimental findings reported by the ALICE collaboration~\cite{ALICE:2022jeh}.
Our predictions are significantly larger than the results from the statistical hadronization model~\cite{Andronic:2019wva,Andronic:2017pug}, which level off at approximately $0.05$ in central collisions. As discussed above, the reason for this is that in our transport approach the regeneration of $\jpsi$'s does not reach its equilibrium limit, while the $\psip$ does, albeit at lower temperatures, recall Figure~\ref{fig:time_evo}.

%

\begin{figure}[!h]
   \centering
   \includegraphics[width=6.8cm]{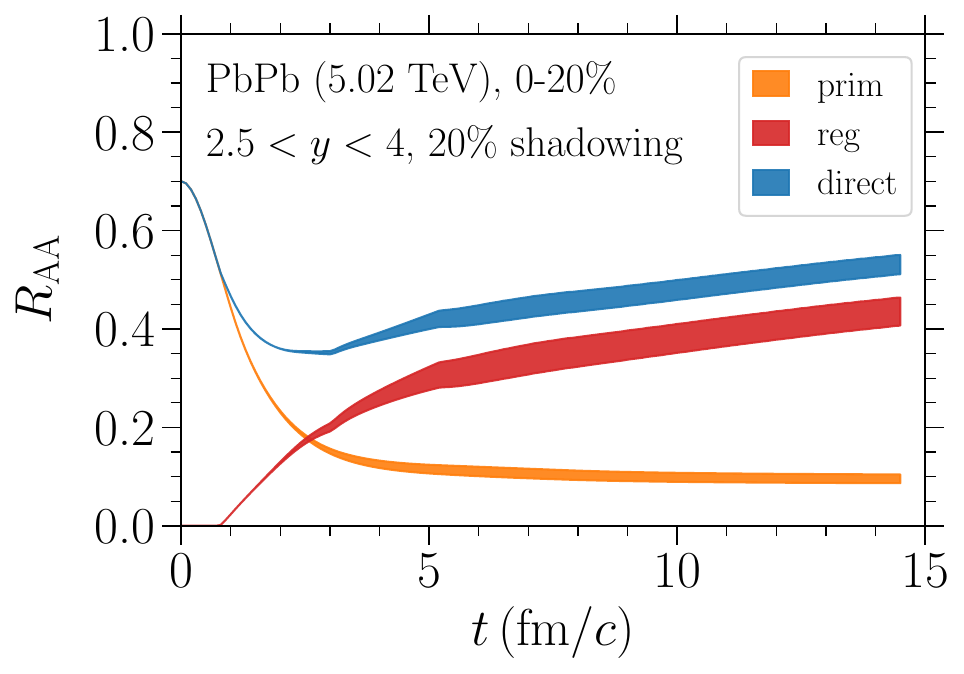}
   \includegraphics[width=6.8cm]{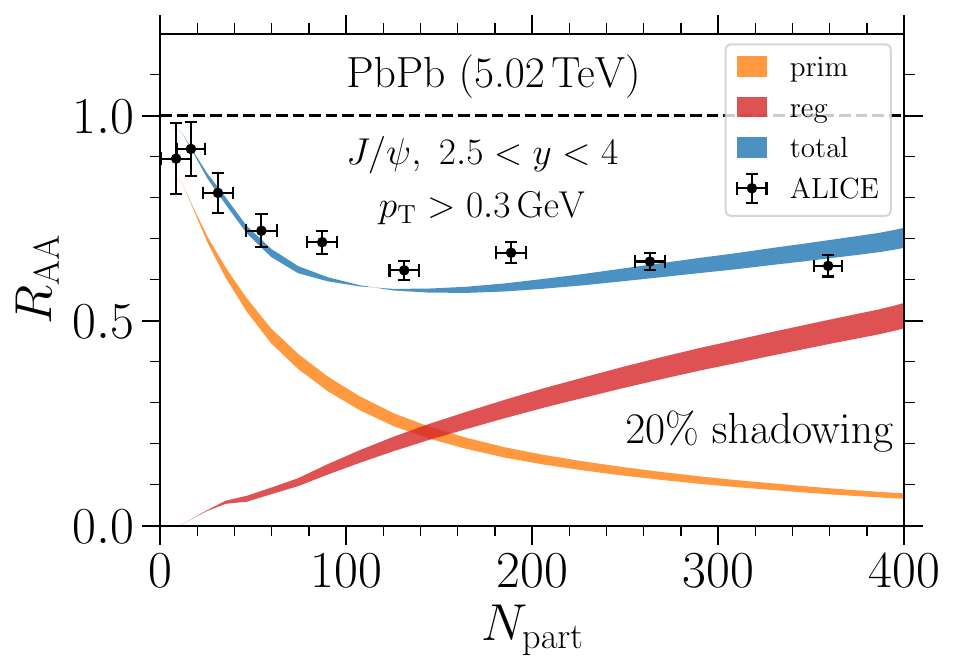}
   \caption{The time evolution (left panel) and centrality dependence (right panel, additionally including $B$ feeddown) of the regenerated (red), primordial (orange), and total (blue) $\jpsi$ production, with the bands illustrating the uncertainty as to whether gluo-dissociation rates are accounted for or not. The same values for shadowing and the $c\bar c$ cross section as in Figure~\ref{fig:time_evo} are used, and the ALICE data are from Ref.~\cite{ALICE:2023gco}.}
\label{fig:gd}
\end{figure}

Next, we return to scrutinizing the impact of the gluo-dissociation processes on our results by incorporating the pertinent rates into the rate equation\footnote{Strictly speaking, this is not following the philosophy of our approach where we adjust the main parameters, \ie, the effective coupling constant, $\alpha_s$, in the quasifree rates as well as the thermal relaxation rate of charm quarks, to match SPS and RHIC data~\cite{Grandchamp:2002wp,Zhao:2010nk}. However, the small impact of gluo-dissociation on our results renders this exercise rather obsolete.}.
In Figure~\ref{fig:gd}, we present results for the time evolution in central collisions (left panel; for direct production) and the centrality dependence (right panel; for inclusive production) of the regenerated, primordial and total $\jpsi$ $\raa$ with and without gluo-dissociation at forward rapidity. Note that a higher rate implies both stronger suppression and increased regeneration. For peripheral and semi-central collisions these two effects essentially compensate each other, while for central collisions, the total $\jpsi$ yield increases by $\sim$6\% due to the enhanced regeneration.

%

Let us also briefly come back to the effect of the updated binding energies as compared to our previous calculations, as mentioned in the first paragraph of Sec.~\ref{ssec_rates}. The impact of the somewhat larger binding energies (most notably around $T$$\simeq$220\,MeV) on regeneration is negligible, as one finds an almost complete compensation between the reduced rates and the larger equilibrium limits. The smaller rates do, however, affect the primordial suppression. The maximal effect for central Pb-Pb collisions amounts to an increase of the primordial contribution by about 0.03 units in terms of the $\raa$.

\section{Transverse-Momentum Spectra}
\label{sec:pt}
In this section, we follow the methodology outlined in Refs.~\cite{Zhao:2010nk} to calculate charmonium $\pT$ spectra utilizing the results from the rate equation.
As discussed toward the end of Sec.~\ref{ssec_trans}, the spectra can be decomposed into primordial and regenerated components according to
\begin{equation}
\frac{\dd N_{\psi}^{\rm PbPb}}{\dd \pT^2}=\frac{\dd N_{\psi}^{\rm prim}}{\dd \pT^2}+\frac{\dd N_{\psi}^{\rm reg}}{\dd \pT^2} \  ,
\end{equation}
with a pertinent nuclear modification factor
\begin{equation}
\raa (\pT) =  \frac{\dd N_{\psi}^{\rm prim}/\dd \pT^2 + \dd N_{\psi}^{\rm reg}/\dd \pT^2}{\Ncoll \dd N_{\psi}^{pp}/\dd \pT^2} \ .
\end{equation}
We solve for the primordial part by employing the Boltzmann equation without the gain term, with initial conditions obtained from pp collisions as specified in Sec.~\ref{ssec_prim}). The yield from regeneration then follows from the difference of the homogeneous solutions and the full rate equation, and we assume its $\pT$ dependence to be given by a thermal-blastwave expression for an average regeneration temperature based on our expansion model (Sec.~\ref{ssec_regen}). This approximation has, of course, its limitations, and we will discuss evidence for that in systematic comparisons to experimental data which will be carried out in Sec.~\ref{ssec_exp}.


\subsection{Initial $\pT$ spectra and their suppression in heavy-ion collisions}
\label{ssec_prim}

%
\begin{figure}[!t]
   \begin{minipage}[b]{1\linewidth}
      \centering
      \includegraphics[width=0.98\textwidth]{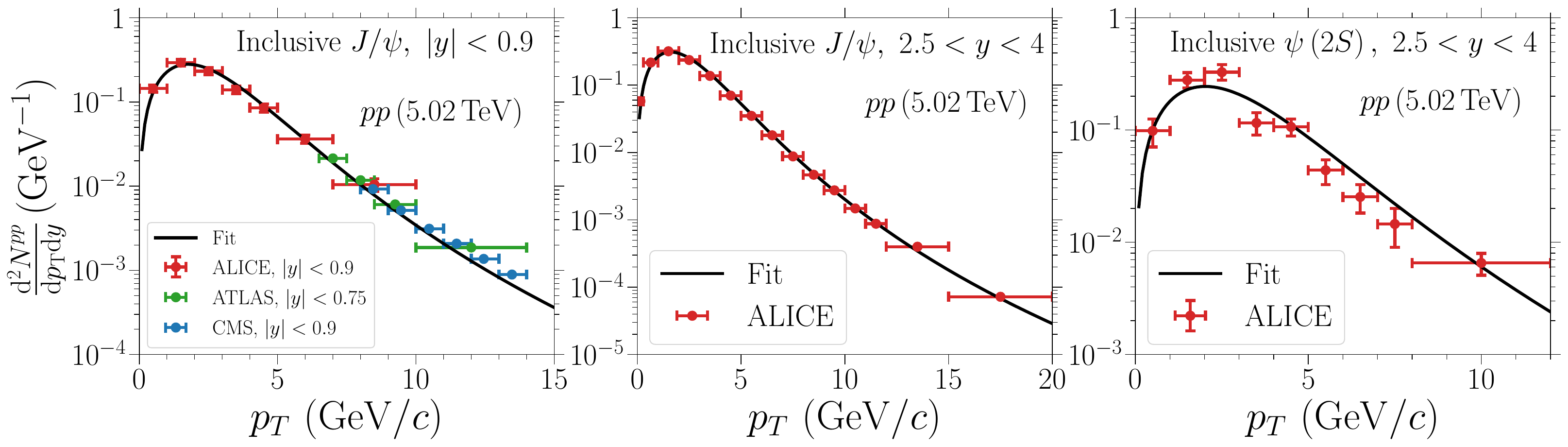}
   \end{minipage}
   \caption{
    Normalized $\pT$ spectra fitted to ALICE data of inclusive $\jpsi$ production in pp collisions at mid-rapidity~\cite{ALICE:2019pid,ATLAS:2017prf,CMS:2017exb} (left panel), forward rapidity~\cite{ALICE:2021qlw} (middle panel), and inclusive $\psip$ at forward rapidity~\cite{ALICE:2021qlw}.
}
  \label{fig:jpsi_pp}
\end{figure}
To construct the initial conditions of charmonia, we first perform fits to their $\pT$ spectra in pp (5.02\,TeV) collisions from the  ALICE~\cite{ALICE:2019pid,ALICE:2021qlw}, ATLAS~\cite{ATLAS:2017prf} and CMS~\cite{CMS:2017exb} collaborations based on the ansatz
\begin{equation}
   \frac{\dd N^{pp}_{\psi}}{\dd \pT^2}=\frac{N}{\left(1+\left({\pT}/{A}\right)^2\right)^n} \ .
   \label{p_T}
\end{equation}
For the three parameters, we obtain $N=0.043$, $A=4.48$ and $n=3.73$ at mid-rapidity and $N=0.052$, $A=4.20$ and $n=3.90$ at forward rapidity for the $\jpsi$ spectra, and $N=0.033$, $A=5.10$ and $n=3.70$ for $\psip$ spectra at forward rapidity, see Figure~\ref{fig:jpsi_pp}.
These spectra, subjected to nuclear-shadowing effects and with  $b$-feeddown subtracted from the inclusive spectra as described in Sec.~\ref{ssec_ini}, serve as initial-momentum distributions for direct production in AA collisions at a given centrality for both forward and mid-rapidity. In addition, we account for the spatial distributions of the initial charmonia, adopting a binary-collision profile obtained from the Glauber model~\cite{Miller:2007ri}, and assume a factorization between spatial and momentum distributions, 
$f_{\psi} (\vec{x}, \vec{p}, \tau_0)=f_{\psi}(\vec{x}) f_{\psi}(\Vec{p})$.  
We are then in a position to solve the $\pT$-dependent Boltzmann equation for the suppression of primordial production for each state, $\psi$, using the momentum-dependent rates discussed in Sec.~\ref{ssec_rates} within our fireball model. This can be done analytically~\cite{Yan:2006ve} resulting in
\begin{equation}
\begin{aligned}
f_{\psi}(\vec{x},\vec{p},\tau)=f_{\psi}(\vec{x}-\vec{v}(\tau-\tau_0),\vec{p},\tau_0)
{\rm e}^{-\int\limits_{\tau_0}^{\tau}\Gamma_{\psi}(\vec{p},T(\tau'))\dd \tau'} \ .
\label{eq:pt_prim}
\end{aligned}
\end{equation}
For inclusive spectra, as usually presented by the ALICE collaboration, we also need to add back the bottom feeddown contribution, which we assume to be conserved at the (integrated) level of 13\% of the $\Ncoll$-scaled yields. However, for its $\pT$ shape in the $\raa$ we need to account for $b$-quark energy loss and the associated redistribution to lower momenta. This effect has been explicitly assessed in Ref.~\cite{He:2021zej} based on microscopic $b$-quark diffusion calculations and turns out to result in an approximately flat $\raa$ for the daughter $\jpsi$ mesons, which we will assume here for both $\jpsi$ and $\psip$.

\subsection{Transverse-momentum spectra from regeneration}
\label{ssec_regen}
Regarding the $\pT$ spectra of the regenerated component, we follow earlier works in Refs.~\cite{Zhao:2011cv} and employ the blastwave model based on our fireball evolution, thereby assuming that charm quarks have reached thermal equilibrium in the QGP. Based on the most recent open-charm hadron phenomenology in URHICs, which suggests $c$-quark relaxation times in the QGP of $\sim$3-4\,fm~\cite{He:2022ywp,Capellino:2022nvf}, and in light of the fireball lifetimes shown in Figure~\ref{fig:temp}, the assumption can be justified for central AA collisions at the LHC, but is probably not quantitatively accurate in semi-central and questionable in peripheral collisions, cf.~also Ref.~\cite{He:2021zej}. For each charmonium state, one has
\begin{equation}
\frac{\dd N^{\rm reg}_\psi}{\dd p_T^2}= N_0(b) m_T\int_{0}^{R}rdrK_1(\frac{m_T \cosh\rho(r)}{T})I_0(\frac{\pT \sinh\rho(r)}{T}) \ , 
\label{blastwave}
\end{equation}
where $m_T=\sqrt{\pT^2+m_\psi^2}$ denotes the transverse mass and $N_0(b)$ normalizes the absolute yield to the result of the rate equation, Eq.~\ref{eq:rate_eq};
$K_1$ and $I_0$ are the modified Bessel functions of the second and first kind, respectively. The radial flow rapidity, $\rho(r)$, is given by $\rho(r)=\tanh^{-1}\left(v_s \frac{r}{R}\right)$, where $R$ is the radius of the fireball and $v_s$ its surface velocity.
We evaluate this expression at an average evolution time when most of the pertinent regeneration yield has built up, \ie, in the middle of the mixed phase for $\jpsi$ and in the hadronic phase at $T=160$\,MeV for $\psip$, independent of centrality (\eg, $\tau$=6.6 and 9.2\,fm/$c$ for central collisions, recall the left panel in Figure~\ref{fig:time_evo}), see also Ref.~\cite{Du:2015wha}.

\begin{figure}[!t]
   \centering
   \includegraphics[width=13cm]{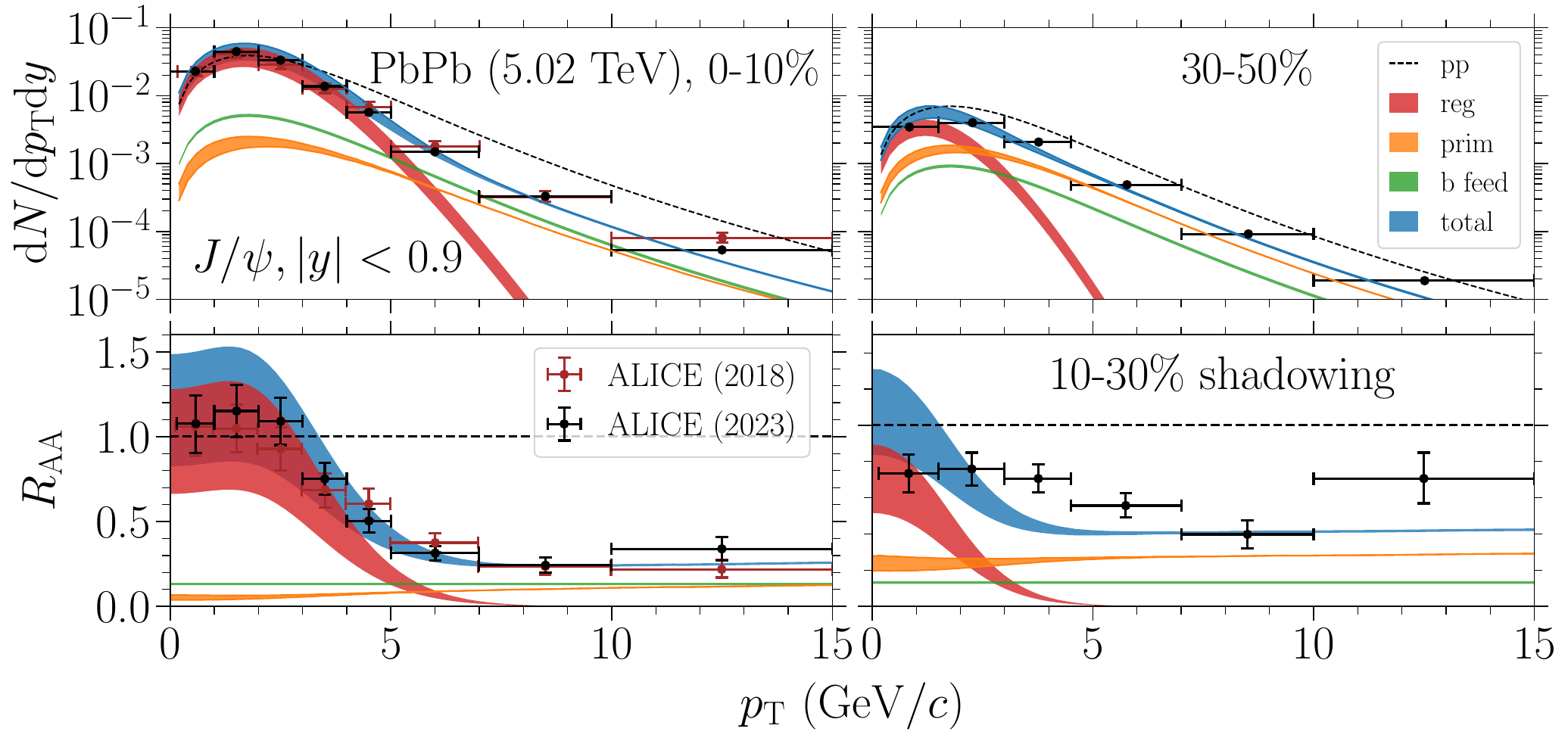}
   \caption{Transverse-momentum spectra (upper panels) and pertinent nuclear-modification factors (lower panels) of inclusive $\jpsi$ production in central (left panels) and semi-central (right panels) Pb-Pb (5.02\,TeV) collisions at mid-rapidity, compared to ALICE data~\cite{Bai:2020svs,ALICE:2023gco}.
    The spectra in pp collisions, scaled by the pertinent binary-collision number, $\Ncoll(b)$, are shown as dashed lines in the upper panels. The bands and the colors of the data have the same meaning as in Figure~\ref{fig:npart}.
    }
    \label{fig:pt_jpsi_mid}
\end{figure}
%
\subsection{Comparison to experimental $\pT$ spectra}
\label{ssec_exp}
%
\begin{figure}[!b]
   \centering
    \includegraphics[width=13cm]{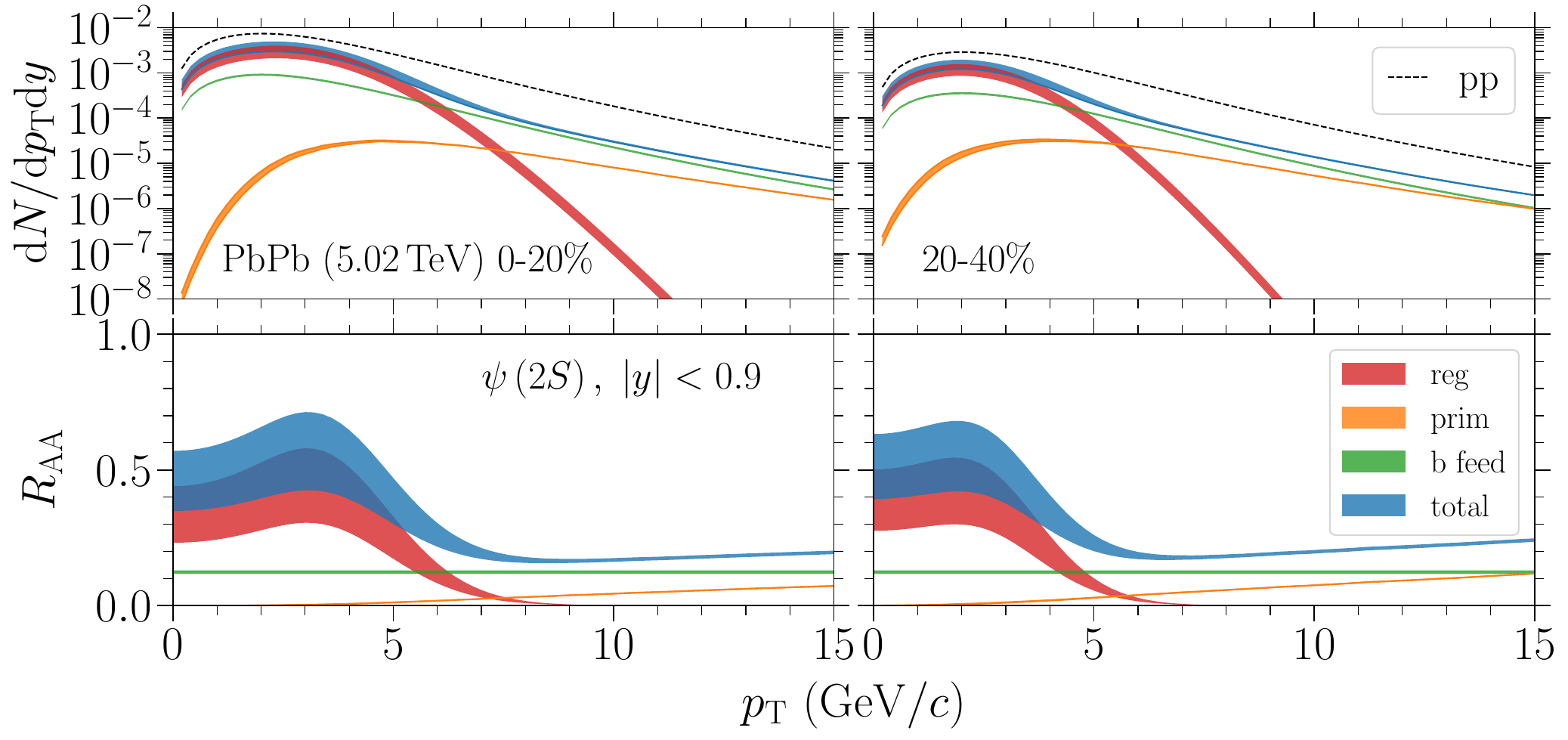}
    \caption{Same as Figure~\ref{fig:pt_jpsi_mid} but for $\psip$ (and without experimental data).
    }
   \label{fig:pt_psip_mid}
\end{figure}
We are now in a position to discuss our results in comparison to experimental data at the LHC, focusing on Pb-Pb(5.02\,TeV) collisions. 
We start with the decomposition of the inclusive $\jpsi$ $\pT$ spectra and their pertinent $\raa$ at mid-rapidity, as shown in   Figure~\ref{fig:pt_jpsi_mid}. 
In 0-10\% central collisions (left panels), we find the well-established features of a strong suppression at high momenta, $\pT \gsim 5\,{\rm GeV}/c$, and a marked rise toward lower $\pT$ due to regeneration, reaching $\raa$ values of one or even larger (mostly depending on the strength of nuclear shadowing that suppresses charm production). The magnitude and shape of this bump, as well as its transition to the rather flat suppression-dominated regime at high $\pT$, are in good agreement with ALICE data, indicating that the blastwave approximation with a collective flow of thermalized charm quarks recombining into $\jpsi$ works well. In 30-50\% semi-central collisions (right panels), both the high-$\pT$ suppression level and the low-$\pT$ recombination bump are less pronounced; however, at the lowest $\pT$, the data tend to be overestimated, while for intermediate $\pT$ around $\sim 5\,{\rm GeV}/c$, the data are underestimated. This discrepancy indicates that the assumption of a thermalized blastwave for the recombining charm quarks is not accurate anymore; indeed, Refs.~\cite{He:2021zej,Du:2022uvj} have shown that employing transported $c$-quark spectra, which do not fully thermalize in semi-central collisions, remedies this discrepancy.  In particular, the crossing between the regeneration and the primordial contribution will be shifted to higher $\pT$, closer to $5\,{\rm GeV}/c$ in semi-central collisions, rather than $3\,{\rm GeV}/c$ as implied by the thermalized blastwave approximation depicted in the right panels of Figure~\ref{fig:pt_jpsi_mid}.

\begin{figure}[!t]
   \centering
      \includegraphics[width=0.98\textwidth]{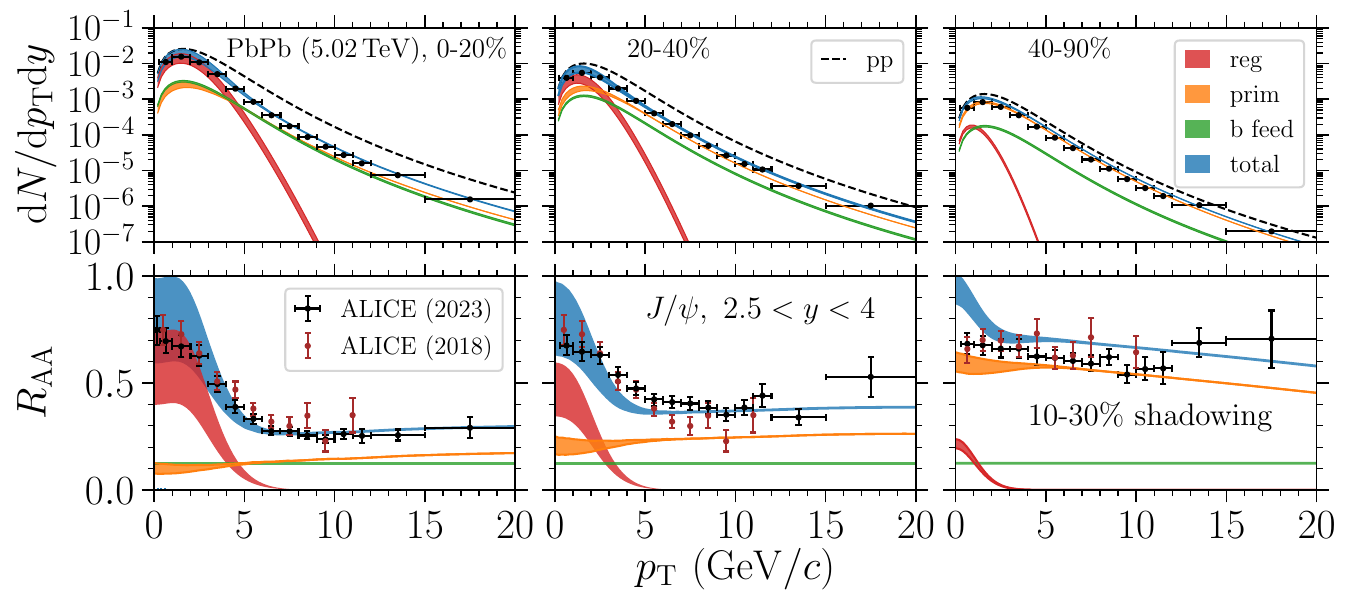}
   \caption{Forward-rapidity $\jpsi$ $\pT$ spectra (upper panels) and $\raa$ (lower panels) compared to ALICE data~\cite{ALICE:2019lga,ALICE:2023gco} for 3 centrality selections.
     Bands and lines have the same meaning as in Figure~\ref{fig:npart}.}
    \label{fig:pt_jpsi_forward}
\end{figure}

Our predictions for the $\pT$ dependence of $\psip$ production at mid-rapidity are shown in Figure~\ref{fig:pt_psip_mid}. At the level of the $\raa$, the regeneration is less prominent for $\psip$ than for $\jpsi$. The later production in the time evolution of the fireball leads to a significant shift of the maximum of the ``flow bump" out to higher $\pT$ compared to $\jpsi$, as a direct consequence of the ``sequential regeneration"~\cite{Du:2015wha}. This effect is also visible when comparing central to semi-central collisions. 
\begin{figure}[!b]
   \centering
   \includegraphics[width=13cm]{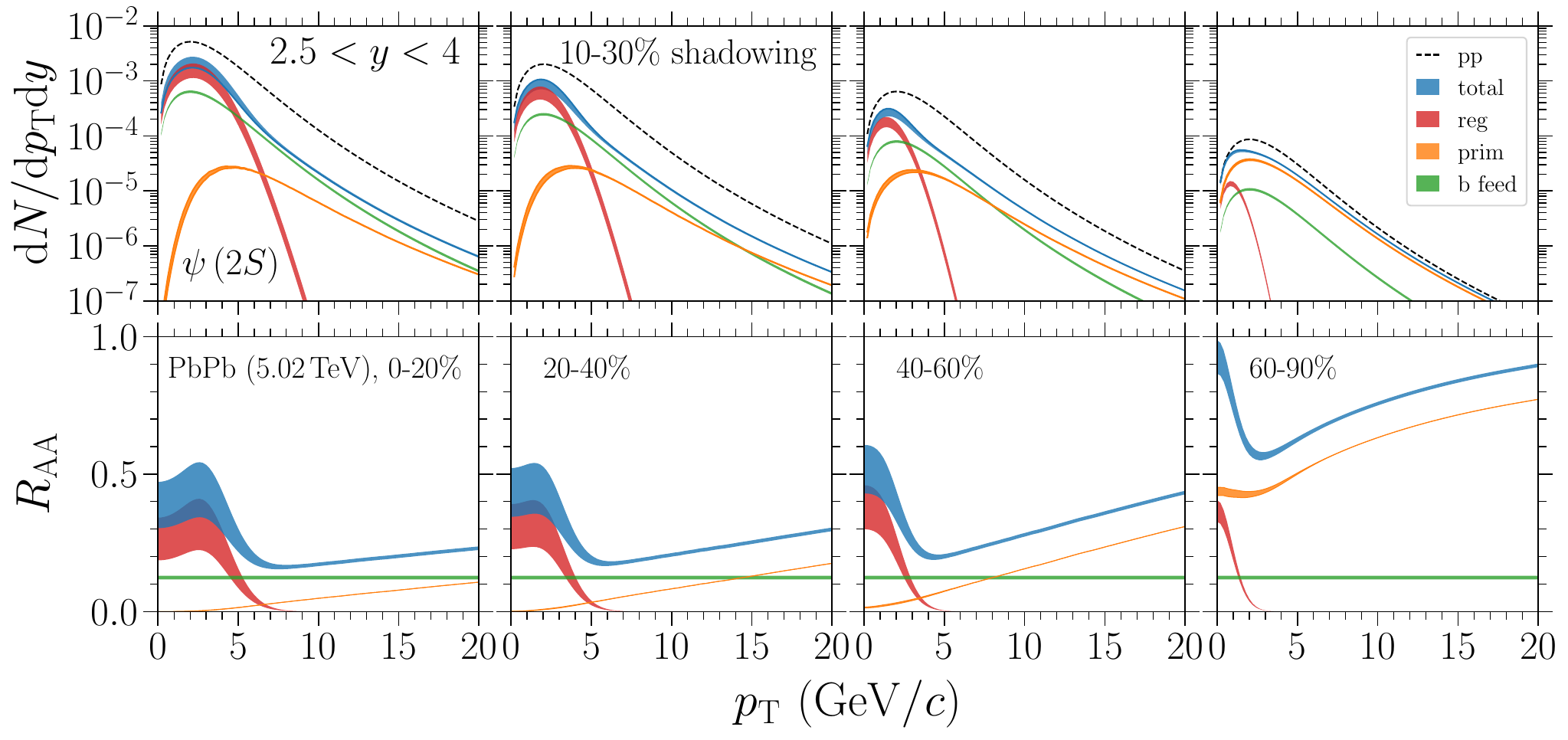}
   \caption{Same as Figure~\ref{fig:pt_jpsi_forward} but for $\psip$ for 4 different centralities.
   }
    \label{fig:pt_psip_forward}
\end{figure}

Next, we turn to forward rapidities where ALICE dimuon data are available. Figure~\ref{fig:pt_jpsi_forward} presents the results for three centrality bins for $\jpsi$ production. The main features of the previously discussed mid-rapidity results persist, including the trend that the regeneration contribution provides a good description of the $\pT$ shape at both low and intermediate values in central collisions. This becomes slightly worse in semi-central collisions, especially for intermediate $\pT$, while in peripheral collisions, the description of the data at low $\pT$ falls apart. This corroborates that the blastwave approximation for the recombined $\jpsi$'s predicts $\pT$ spectra that are increasingly too soft in more peripheral collisions, albeit the contribution to the integrated yield  becomes rather small.

\begin{figure}[t]
\centering
   \includegraphics[width=9.5cm]{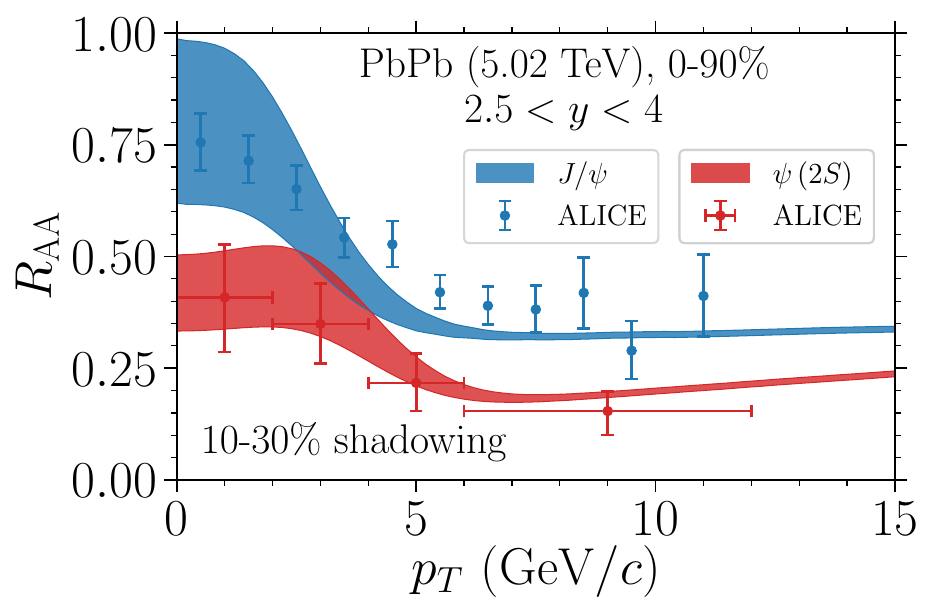}
 \caption{Nuclear-modification factor as a function of $\pT$ for inclusive $\jpsi$ (blue) and $\psip$ (red) production at forward rapidity in 0-90\% Pb-Pb(5.02\,TeV) collisions, compared to ALICE data~\cite{ALICE:2022jeh}.}
     \label{fig:psi_pt}
   \end{figure}
The $\pT$ dependence of $\psip$ production at forward rapidity is summarized in Figure~\ref{fig:pt_psip_forward}. Again, these spectra, along with their corresponding $\raa$'s, share essentially the same features as observed at mid-rapidity. However, the effects are slightly less pronounced, in terms of both the suppression (because of a slightly less hot and shorter-lived fireball due to the lower charged-particle multiplicity at forward rapidity), and the smaller regeneration contribution (due to the smaller charm cross section). Additionally, we expect enhancement at low-$\pT$ for the more peripheral centrality bins in our calculations to overestimate future data. Nevertheless, the regeneration maxima show a systematic shift to higher momenta, due to the increasing transverse flow in more central collisions.

\begin{figure}[!b]
   \includegraphics[width=6.8cm]{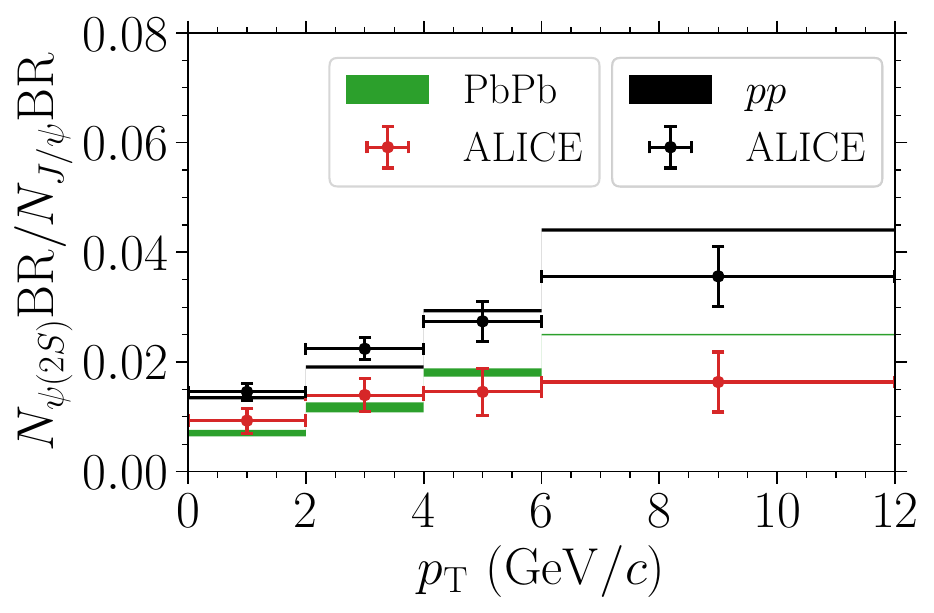}
   \includegraphics[width=6.8cm]{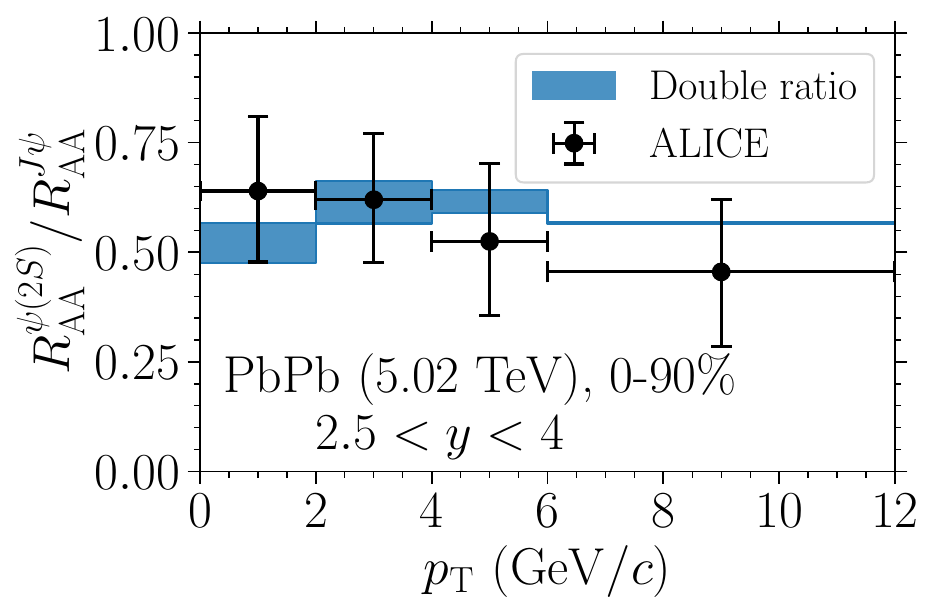}
 \caption{Transverse-momentum dependent $\psip$ over $\jpsi$ ratio (left panel) and their $\raa$ double ratio (right panel) at forward rapidity compared to ALICE data~\cite{ALICE:2022jeh}.}
     \label{fig:double_ratio}
   \end{figure}
\begin{figure}[!t]
   \centering
   \includegraphics[width=13cm]{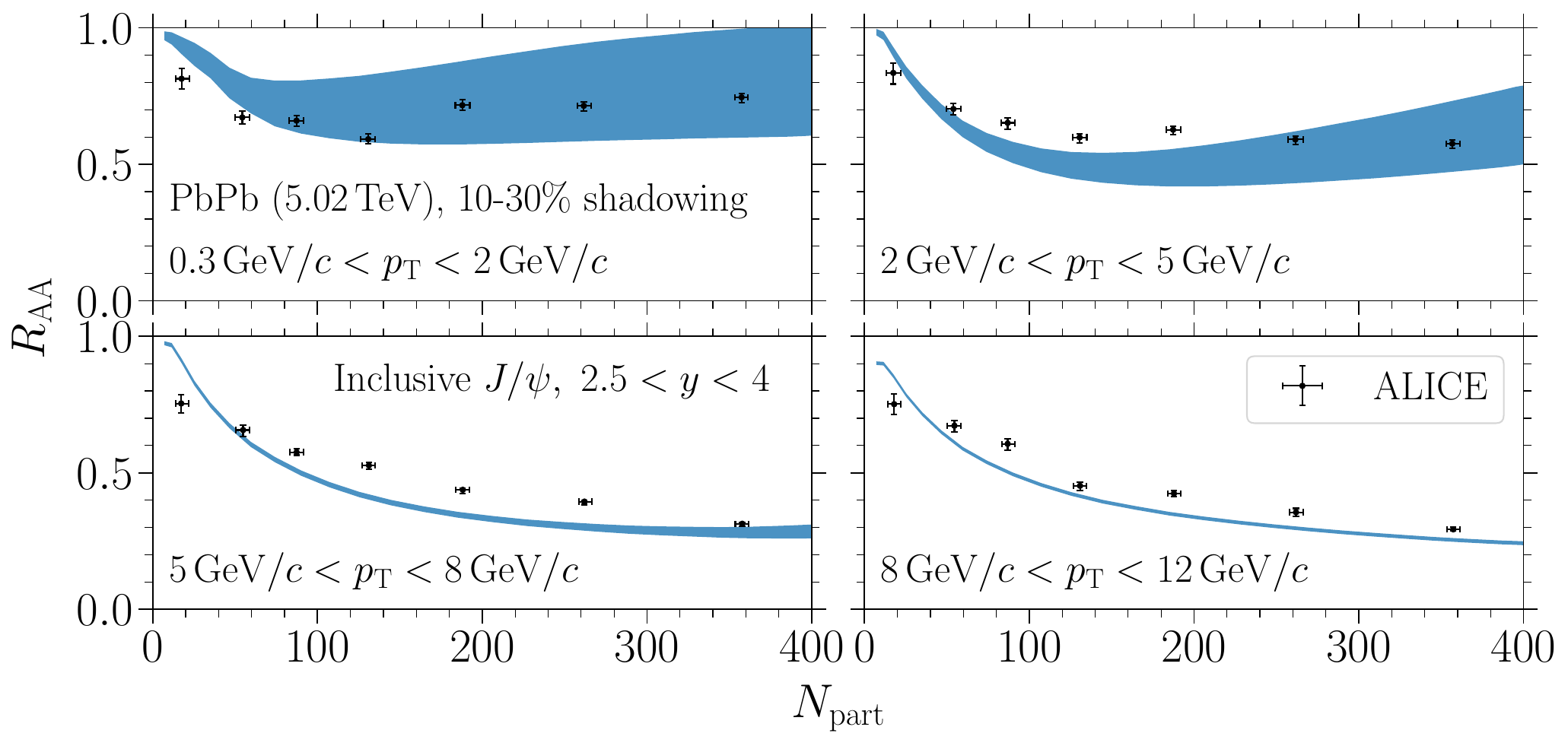}
        \caption{Inclusive $\jpsi$ nuclear-modification factor as a function of $\npart$ for Pb-Pb collisions at 5.02\,TeV for four $\pT$ intervals, compared with data~\cite{ALICE:2019lga}}
     \label{fig:npart_bins}
\end{figure}
Experimental data for the $\pT$ dependence of $\psip$ production in 0-90\% (essentially ``minimum bias") Pb-Pb collisions have recently become available in Ref.~\cite{ALICE:2022jeh}, which also contains our original predictions (consistent with our calculations presented herein). The pertinent $\raa(\pT)$ is illustrated in Figure~\ref{fig:psi_pt}, alongside the $\jpsi$ results. Note that the 0-90\% centrality selection exhibits a rather strong bias toward central collisions, due to the approximate $\Ncoll$ scaling of the hard-produced $c\bar c$ pairs (both open and hidden). Consequently, there is only a slight indication that the blastwave approximation for regenerated $\jpsi$ is inaccurate, as it under-predicts the data around $\pT\simeq 5\,{\rm GeV}/c$. For the $\psip$, the predictions work out well, attesting to the predictive power of our transport framework. Recall that $\psip$ recombination occurs later in the fireball's evolution than that of $\jpsi$, allowing more time for the open-charm particles to relax toward equilibrium; one furthermore finds the ``flow bump" to be moved out to higher $\pT$ compared to the $\jpsi$, which is not conclusive from the data (yet).

The $\pT$ dependent ratio of $\psip$ to $\jpsi$ yields at forward rapidity, as depicted in the left panel of Figure~\ref{fig:double_ratio}, also aligns reasonably well with the ALICE data~\cite{ALICE:2022jeh}. 
The increasing trend with $\pT$ from our fits to pp data tends to overestimate the data for the ratio at the highest $\pT$. This discrepancy also migrates into the AA result (which may be partly due to the underestimation of the $\jpsi$ production at intermediate $\pT$).  
Indeed, the agreement is better when dividing out the pp reference spectra in the $\raa$ double ratio, shown in the right panel of Figure~\ref{fig:double_ratio}.
Here, the slight deficit in the $\jpsi$ yield at intermediate $\pT$ shows up as a mild maximum structure which is not observed in the data at this point. In principle, such a maximum could be another signature of sequential regeneration in central collisions, when the open-charm spectra are close to thermal equilibrium.  

We further investigate the $\raa(\npart)$ for inclusive $\jpsi$ production at forward rapidity, presented in Figure~\ref{fig:pt_jpsi_forward} but binned into different $\pT$ intervals. The corresponding results, alongside ALICE data~\cite{ALICE:2019lga}, are depicted in Figure~\ref{fig:npart_bins}. The $\pT$-dependent shadowing effect, concentrated at low $\pT$, introduces an uncertainty in the $\raa$ which is most pronounced at low $\pT$ but diminishes at higher $\pT$.
In peripheral collisions, the nuclear modification factor is primarily influenced by the primordial contributions across all $\pT$ regions, approaching 1 for small $\npart$. Conversely, central collisions the scenario changes. For the lower $\pT$ bins, regeneration processes predominantly contribute to the observed $\jpsi$ production, leading to an increasing $\raa$, while at higher $\pT$, the contributions shift towards non-prompt and primordial $\jpsi$ production mechanisms, rendering a characteristic decrease in $\raa$ values with increasing $\npart$.

\begin{figure}[!t]
      \centering
      \includegraphics[width=13cm]{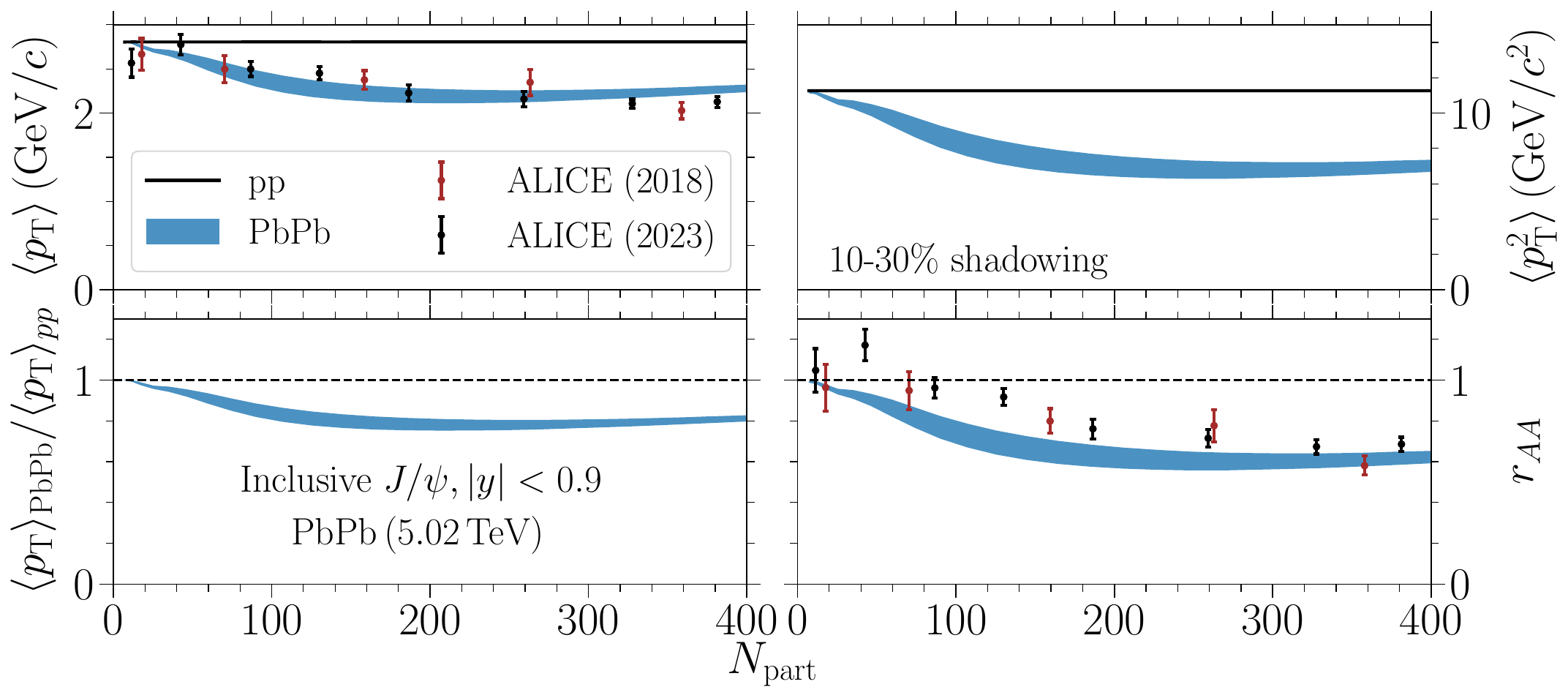}
      \caption{$\langle \pT \rangle$ (upper left) and $\langle \pT^2\rangle$ (upper right) of $\jpsi$ in 5.02 TeV pp and Pb-Pb collisions at mid-rapidity. Their ratios relative to the pp results are shown in the lower panels; ALICE data are from Refs.~\cite{ALICE:2022jeh,ALICE:2019nrq}.}
     \label{fig:mean_pt}
   \end{figure}

Finally, we attend to the centrality dependence of the average charmonium momentum, $\langle\pT\rangle$, and its square, $\langle\pT^2\rangle$. We calculate a ``small" nuclear-modification factor" defined in Ref.~\cite{Zhou:2014kka}, normalized to the values in pp and Pb-Pb collisions, 
\begin{equation}
   r_{\rm AA} = \frac{\langle \pT^2 \rangle_{\rm PbPb}}{\langle \pT^2 \rangle_{pp}} \ ;
\end{equation}
the results are shown in Figure~\ref{fig:mean_pt}.
In central collisions, the predominance of regenerated $\jpsi$'s, which typically exhibit a softer spectrum, leads to a reduction in both the transverse momentum and its square, causing their $r_{\rm AA}$'s to fall below 1. This effect is more pronounced for the latter and also shows a more gradual fall-off with centrality. 
 Our model calculations describe the centrality dependence of  $\langle\pT\rangle$ rather well, yet slightly underestimate the $\langle\pT^2\rangle$ data in semi-central collisions. Again, this can be traced back to a recombination contribution that is too soft in semi-central collisions, most likely since charm quarks do not achieve full thermal equilibrium, which can be remedied by employing explicitly transported charm-quark distributions~\cite{He:2021zej,Du:2022uvj}.

\section{Conclusions}
\label{sec:concl}
We have investigated the production of charmonia in ultra-relativistic heavy-ion collisions using a previously constructed semi-classical transport approach that satisfies detailed balance and incorporates gradual quarkonium dissociation utilizing reaction rates based on in-medium binding energies and heavy-quark masses.
Compared to our previous studies, notable updates include revised in-medium binding energies guided by recent $T$-matrix computations, state-of-the-art charm production cross sections from experiment and their $\pT$ dependent shadowing. Our focus has been on charmonium kinetics in 5.02\,TeV Pb-Pb collisions, and specifically on predictions for $\psip$ observables that recently became available at the LHC. An important role is being played by the mechanism of ``sequential regeneration", where regeneration processes for the $\psip$ are operative at lower temperatures than for the $\jpsi$, with significant contributions also from the hadronic phase. While the total $\jpsi$ yield is close to its chemical equilibrium values in the QGP phase of central Pb-Pb collisions (around temperatures of $\sim$250\,MeV), the $\psip$ yields chemically equilibrate later, at temperatures of $\sim$160\,MeV. This delay has significant consequences for observables, most notably a $\psip/\jpsi$ ratio above the equilibrium values at any given temperature and a shift of the ``flow bump" in the nuclear modification factor to higher momenta for $\psip$ than for $\jpsi$. The former has been confirmed by experiment, while the latter is a more subtle effect that the data are not (yet) sensitive to. Furthermore, we have also re-confirmed the rather negligible effect of gluo-dissociation on the reaction rates and highlighted limitations of our blastwave approximation for the $\pT$ spectra of the regenerated charmonia, which leads to an overestimation of the low-$\pT$ yields in peripheral Pb-Pb collisions. On the other hand, in central collisions the $\pT$-dependent $\raa$'s align well with the ALICE data, indicating that the charm-quark spectra at low and intermediate momenta are near local thermal equilibrium.
Improvements by implementing the full kinetics of charm-quark diffusion have already been worked out for specific cases (and enabled, \eg, a resolution of the so-called $\jpsi$ $v_2$ puzzle~\cite{He:2021zej}), but are still awaiting systematic applications to the full available data samples. Further objectives of future developments are the implementation of nonperturbative matrix elements for the quasifree processes, and a realistic implementation of quantum transport for charmonia~\cite{Andronic:2024oxz} that, in particular, can cope with regeneration reactions in the presence of multiple charm-anticharm quark pairs. Work in all these directions is in progress.


\acknowledgments 

This work has been supported by the U.S. National Science Foundation under grant nos. PHY-1913286 and PHY-2209335, by the TAMU Cyclotron Institute's Research Development (CIRD) program, and by the U.S. Department of Energy, Office of Science, Office of Nuclear Physics through the Topical Collaboration in Nuclear Theory on \textit{Heavy-Flavor Theory (HEFTY) for QCD Matter} under award no.~DE-SC0023547.

\begin{adjustwidth}{-\extralength}{0cm}

\reftitle{References}


\bibliography{refcnew}

\PublishersNote{}
\end{adjustwidth}

\end{document}